%% file: qz13tev.tex
\pdfoutput=1
\newcommand*{\ATLASLATEXPATH}{latex/}
\documentclass[UKenglish,cernpreprint,texlive=2016]{\ATLASLATEXPATH atlasdoc}
\usepackage{\ATLASLATEXPATH atlaspackage}
\usepackage{\ATLASLATEXPATH atlasbiblatex}
\usepackage{slashed}
\usepackage{bm}
\usepackage{float}
\usepackage{multirow}
\usepackage{\ATLASLATEXPATH atlasphysics}
\graphicspath{{logos/}{figures/}}
\usepackage{qz13tev-defs}
\input{qz13tev-metadata.tex}
\hypersetup{pdftitle={ATLAS document},pdfauthor={The ATLAS Collaboration}}
\begin{document}

\maketitle

\input{qz13tev-intro.tex}
\input{qz13tev-detector.tex}
\input{qz13tev-samples.tex}
\input{qz13tev-reconstruction.tex}
\input{qz13tev-selection.tex}
\input{qz13tev-backgrounds.tex}
\input{qz13tev-systematics.tex}
\input{qz13tev-limits.tex}
\input{qz13tev-conclusions.tex}

\section*{Acknowledgements}
\input{acknowledgements/Acknowledgements}

\printbibliography

\clearpage \input{atlas_authlist}
\end{document}

%% file: qz13tev-metadata.tex
\AtlasTitle{Search for flavour-changing neutral current top-quark decays $t\to qZ$ in proton--proton collisions at $\sqrt{s}=13$ TeV with the ATLAS detector}

\author{The ATLAS Collaboration}

\PreprintIdNumber{CERN-EP-2018-018}

\AtlasJournal{JHEP}
\AtlasJournalRef{JHEP 07 (2018) 176}
\AtlasDOI{DOI:10.1007/JHEP07(2018)176}

\AtlasAbstract{
A search for flavour-changing neutral-current processes in top-quark
decays is presented.  Data collected with the ATLAS detector from proton--proton collisions at
the Large Hadron Collider at a centre-of-mass energy of $\sqrt{s}=13$ TeV,
corresponding to an integrated luminosity of 36.1 fb$^{-1}$, are
analysed.  The search is performed using top-quark pair
events, with one top quark decaying through the $t\rightarrow qZ$ $(q
= u, c)$ flavour-changing neutral-current channel, and the other
through the dominant Standard Model mode $t\rightarrow bW$.  Only
$Z$ boson decays into charged leptons and leptonic $W$ boson
decays are considered as signal.  Consequently, the final-state
topology is characterized by the presence of three isolated charged
leptons (electrons or muons), at least two jets, one of the jets originating from a
$b$-quark, and missing transverse momentum from the undetected neutrino.
The data are consistent with Standard Model background
contributions, and at 95\% confidence level the search sets observed (expected) upper limits 
of $1.7\times10^{-4}$ ($2.4\times10^{-4}$) on the $t\to uZ$ branching ratio 
and $2.4\times10^{-4}$ ($3.2\times10^{-4}$) on the $t\to cZ$ branching ratio,
constituting the most stringent limits to date.
}

%% file: qz13tev-intro.tex
\section{Introduction\label{sec:intro}}

The top quark is the heaviest elementary particle known, with a mass 
$m_t=173.3\pm0.8$~\GeV~\cite{ATLAS-CONF-2014-008}.
In the Standard Model of particle physics (SM), it decays almost 
exclusively into $bW$ while flavour-changing neutral current~(FCNC) decays such as $t\to qZ$ are 
forbidden at tree level. FCNC decays occur at one-loop level but are strongly
suppressed by the GIM mechanism~\cite{Glashow:1970gm} with a suppression factor of 14
orders of magnitude relative to the dominant decay 
mode~\cite{AguilarSaavedra:2004wm}. However, several SM extensions predict 
higher branching ratios for top-quark FCNC decays. Examples of such 
extensions are the quark-singlet model (QS)~\cite{AguilarSaavedra:2002kr}, 
the two-Higgs-doublet model with~(FC~2HDM) or without~(2HDM) flavour 
conservation~\cite{Atwood:1996vj}, the Minimal Supersymmetric Standard Model 
(MSSM)~\cite{Cao:2007dk}, the MSSM with R-parity 
violation~(RPV SUSY)~\cite{Yang:1997dk}, models with warped 
extra dimensions~(RS)~\cite{Agashe:2006wa}, or extended mirror fermion 
models~(EMF)~\cite{Hung:2017tts}. Reference~\cite{Agashe:2013hma} gives a comprehensive
review of the various extensions of the SM that have been proposed. 
Table~\ref{tab:intro:models} provides the maximum
values for the branching ratios $\mathcal{B}$(\tqz) predicted by these models and 
compares them to the value predicted by the SM.

Experimental limits on the FCNC branching ratio $\mathcal{B}$(\tqz) were established by experiments at 
the Large Electron--Positron collider~\cite{Heister:2002xv, 
Abdallah:2003wf, Abbiendi:2001wk, Achard:2002vv, LEP-Exotica-WG-2001-01}, 
HERA~\cite{Abramowicz:2011tv}, the Tevatron~\cite{Aaltonen:2008ac,Abazov:2011qf},
and the Large Hadron Collider~(LHC)~\cite{TOPQ-2011-22, TOPQ-2014-08, 
CMS-TOP-12-037, CMS-TOP-12-039}. 
Before the results reported here, the most stringent limits were
${\mathcal{B}}(\tuz) < 2.2\times 10^{-4}$ and ${\mathcal{B}}(\tcz) < 4.9\times 10^{-4}$ at 95\% 
confidence level (CL), both set by the CMS Collaboration~\cite{CMS-TOP-12-039} using data collected at 
$\sqrt{s}=8$~\TeV. For the same centre-of-mass energy, the ATLAS 
Collaboration derived a limit of ${\mathcal{B}}(\tqz) < 7\times 10^{-4}$~\cite{TOPQ-2014-08}.
ATLAS results obtained at $\sqrt{s}=7$~\TeV\ are also available~\cite{TOPQ-2011-22}.

This analysis is a search for the FCNC decay
\tqz\ in top-quark--top-antiquark (\ttbar) events, in \luminosity of data collected 
at $\sqrt{s}=13$~\TeV, where one top quark decays through the
FCNC mode and the other through the dominant SM mode ($t\to bW$). 
Only $Z$ boson decays into charged leptons and leptonic $W$ boson decays 
are considered. The final-state topology is thus characterized by the 
presence of three isolated charged leptons,\footnote{In this article, lepton is 
used to denote an electron or muon, including those coming from leptonic $\tau$-lepton
decays.} at least two jets with exactly one being tagged as a jet containing $b$-hadrons, 
and missing transverse momentum from the 
undetected neutrino. The main sources of background events containing three 
prompt leptons are diboson, \ttz, and $tZ$ production.
Events with two or fewer prompt leptons and 
additional non-prompt\footnote{Prompt leptons are leptons from the decay of $W$ or $Z$ bosons, either
directly or through an intermediate $\tau\to \ell\nu\nu$ decay, or from the
semileptonic decay of top quarks.} leptons are also sources of background.
Besides the signal region, control 
regions are defined to constrain the main backgrounds. 
Results are obtained using a binned likelihood fit to kinematic distributions in the signal and control regions.

The article is organized as follows. A brief description of the ATLAS detector is given in 
Section~\ref{sec:detector}. The collected data samples and the simulations of 
signal and SM background processes are described in 
Section~\ref{sec:samples}. Section~\ref{sec:reco} presents the object 
definitions, while the event analysis and kinematic reconstruction are 
explained in Section~\ref{sec:selection}. Background evaluation and sources 
of systematic uncertainty are described in Sections~\ref{sec:backgrounds} 
and~\ref{sec:systematics}. Results are presented in Section~\ref{sec:limits}, 
and conclusions are drawn in Section~\ref{sec:conclusions}.

%% file: qz13tev-detector.tex
\section{ATLAS detector and data samples\label{sec:detector}}

The ATLAS experiment~\cite{PERF-2007-01} is a multi-purpose particle physics detector consisting 
of several subdetector systems, which almost fully cover the solid 
angle\footnote{\label{dr}ATLAS uses a right-handed coordinate system with its 
origin at the nominal interaction point in the centre of the detector and the 
$z$-axis along the beam pipe. The $x$-axis points from the interaction point 
to the centre of the LHC ring, and the $y$-axis points upward. Cylindrical 
coordinates $(r,\phi)$ are used in the transverse plane, $\phi$ being the 
azimuthal angle around the beam pipe. The pseudorapidity is defined in terms 
of the polar angle $\theta$ as $\eta=-\ln\tan(\theta/2)$. The $\Delta R$ 
distance is defined as $\Delta R=\sqrt{(\Delta\eta)^2+(\Delta\phi)^2}$.} 
around the interaction point. It is composed of an inner tracking system 
close to the interaction point and immersed in a 2~T axial magnetic field 
produced by a thin superconducting solenoid. A lead/liquid-argon (LAr) 
electromagnetic calorimeter, a steel/scintillator-tile hadronic calorimeter, 
copper/LAr hadronic endcap calorimeters, copper/LAr and tungsten/LAr forward calorimeters, 
and a muon spectrometer with three 
superconducting magnets, each one with eight toroid coils, complete the detector. 
A new innermost silicon pixel layer was added to the inner detector 
after Run 1~\cite{ATLAS:ibl,ATLAS:ibl2}.
The combination of all these systems provides charged-particle momentum 
measurements, together with efficient and precise lepton and photon 
identification in the pseudorapidity range $|\eta| < 2.5$. Energy deposits 
over the full coverage of the calorimeters, $|\eta| < 4.9$, are used to 
reconstruct jets and missing transverse momentum. A 
two-level trigger system is used to select interesting 
events~\cite{TRIG-2016-01}. The first level is implemented with 
custom hardware and uses a subset of detector information to reduce the event 
rate. It is followed by a software-based trigger level to reduce the event 
rate to approximately 1~kHz. 

In this analysis, the combined 2015 and 2016 datasets from proton--proton ($pp$) 
collisions at $\sqrt{s}=13$~\TeV\ 
corresponding to an
integrated luminosity of  \luminosity
are used. Analysed events are selected by 
either a single-electron or a single-muon trigger.
Triggers with different transverse-momentum 
thresholds are used to increase the overall efficiency. The triggers 
using a low electron transverse momentum~($\pt^e$) or muon transverse momentum~($\pt^\mu$) threshold ($\pt^e>24$~\GeV\ or 
$\pt^\mu>20$~\GeV\ for 2015 data and $\pt^{e,\mu}>26$~\GeV\ for 2016 data) 
also have isolation requirements. At high \pt\, the isolation 
requirements incur small efficiency losses which are 
recovered by higher-threshold triggers ($\pt^e>60$~\GeV, $\pt^e>120$~\GeV, or 
$\pt^\mu>50$~\GeV\ for 2015 data and $\pt^e>60$~\GeV, $\pt^e>140$~\GeV, or 
$\pt^\mu>50$~\GeV\ for 2016 data) without isolation requirements.

\begin{table}[t]
\caption{Maximum allowed FCNC \tqz\  $(q = u, c)$ branching ratios predicted by several 
models~\cite{AguilarSaavedra:2004wm, AguilarSaavedra:2002kr, Atwood:1996vj, 
Cao:2007dk, Yang:1997dk, Agashe:2006wa, Agashe:2013hma, Hung:2017tts}.}
\label{tab:intro:models}
\begin{center}
\begin{tabular}{lcccccccc}
  \hline
  Model:	& SM		& QS		& 
  2HDM		& FC 2HDM	& MSSM		& 
  RPV SUSY	& RS	& EMF		\\
  \hline\\[-1em]
  $\mathcal{B}$(\tqz):	& $10^{-14}$	& $10^{-4}$	& $10^{-6}$	& $10^{-10}$	& $10^{-7}$	& $10^{-6}$		& $10^{-5\phantom{0}}$		& $10^{-6}$\\
  \hline
\end{tabular}
\end{center}
\end{table}

%% file: qz13tev-samples.tex
\section{Signal and background simulation samples\label{sec:samples}}

In $pp$ collisions at a centre-of-mass 
energy of $\sqrt{s} = 13~\TeV$ at the LHC, top quarks are produced according to the SM mainly in
\ttbar pairs with a predicted cross section of 
$\sigma_{t\bar{t}}=0.83\pm0.05$ nb ~\cite{Cacciari:2011hy,Baernreuther:2012ws,Czakon:2012zr,Czakon:2012pz,Czakon:2013goa,Czakon:2011xx}
for the top-quark mass value of $172.5~\GeV$ used to simulate events 
as described in the following paragraphs;
the uncertainty includes contributions from uncertainties in the factorization and
renormalization scales, the parton distribution functions (PDF), the strong coupling
$\alpha_{\mathrm S}$, and the top-quark mass. 
The cross section is calculated at next-to-next-to-leading order~(NNLO) 
in QCD including resummation of next-to-next-to-leading logarithmic 
soft gluon terms with {\ttf Top++} 
2.0. 
The effects of PDF and $\alpha_{\mathrm S}$ uncertainties 
are calculated using the PDF4LHC prescription~\cite{Botje:2011sn} with the 
MSTW 2008 68\% CL NNLO~\cite{Martin:2009iq,Martin:2009bu}, CT10 
NNLO~\cite{Lai:2010vv,Gao:2013xoa} and NNPDF 2.3 5f FFN~\cite{Ball:2012cx} 
PDF sets and are added in quadrature to those from the renormalization and factorization 
scale uncertainties.

The next-to-leading-order~(NLO) simulation of signal events was performed with the event generator 
{\ttf MG5\_aMC@NLO}~\cite{Alwall:2014hca} interfaced to {\ttf Pythia8}~\cite{Sjostrand:2014zea} with 
the {\ttf A14}~\cite{ATL-PHYS-PUB-2014-021} set of tuned parameters and the {\ttf NNPDF2.3LO} PDF 
set~\cite{Ball:2012cx}. Dynamic factorization and renormalization scales were used.
The factorization and renormalization scales were set equal to
\( \sqrt{m_{t}^2 + (p_{\mathrm{T},t}^2 + p_{\mathrm{T},\bar{t}}^2)/2 \:}\)
where $ p_{\mathrm{T},t}^{\phantom{2}} $ ($ p_{\mathrm{T},\bar{t}}^{\phantom{2}}$) is the transverse momentum of 
the top quark (top antiquark).
For the matrix element, the PDF set {\ttf 
NNPDF3.0NLO}~\cite{Ball:2014uwa} was used. 
For the top-quark FCNC decay, the effects of new physics at 
an energy scale $\Lambda$ were included by adding dimension-six effective terms to the SM 
Lagrangian. No differences between the kinematical distributions 
from the $bWuZ$ and $bWcZ$ processes are observed.
Due to the different 
$b$-tagging mistag rates for $u$- and $c$-quarks, the signal 
efficiencies differ after applying requirements on the $b$-tagged jet 
multiplicity.
Hence limits on $\mathcal{B}$(\tqz) are set separately for $q = u , c$. 
Only decays of the $W$ and $Z$ bosons with charged 
leptons were 
generated at the matrix-element level ($Z\to e^+e^-$, $\mu^+\mu^-$, or $\tau^+\tau^-$ 
and $W\to e\nu$, $\mu\nu$, or $\tau\nu$).

Several SM processes have final-state topologies similar to the signal, with 
at least three prompt charged leptons, especially dibosons ($WZ$ and $ZZ$), 
\ttv\ ($V$ is \Wboson or \Zboson), \ttbar $WW$, \tth, gluon--gluon fusion ($gg$F) $H$,
vector-boson fusion (VBF) $H$, $VH$, $tZ$, 
$WtZ$, $ttt(t)$, and triboson ($WWW$, $ZWW$ and $ZZZ$) production. 
The theoretical estimates for these backgrounds are further
constrained by the simultaneous fit to the signal and control regions
described below. Events with non-prompt leptons or events in which at least one jet is misidentified as an 
isolated charged lepton (labelled as non-prompt leptons throughout this article) 
can also fulfil the event selection requirements. These events, typically 
$Z+$jets (including $\gamma$ emission), 
\ttbar, and single-top ($Wt$), are estimated with the semi-data-driven method explained
in Section~\ref{sec:backgrounds}, which also 
uses simulated samples which for the $Z+$jets events include $Z$ production in association with heavy-flavour quarks.

Table~\ref{tab:samples} summarizes information about the generators, 
parton shower, and PDFs used to simulate the 
different event samples considered in the analysis. 
The associated production of a
\ttbar pair with one vector boson was generated at NLO
with $\texttt{MG5\_aMC@NLO}$ interfaced to
$\texttt{Pythia8}$ with the $\texttt{A14}$ set of tuned parameters and the
$\texttt{NNPDF2.3LO}$ PDF set.
The $t\bar{t}Z$ and $t\bar{t}W$ samples were normalized to the NLO QCD+electroweak cross-section 
calculation using a fixed scale ($m_t+m_V/2$)~\cite{deFlorian:2016spz}. 
In the case of the $t\bar{t}Z$ sample with the $Z\to\ell^+\ell^-$ decay mode, the 
$Z/\gamma^{*}$ interference was included with the criterion $m_{\ell\ell}>5$~\GeV\ 
applied. 
The ~$t$-channel production of a single top quark in association with a $Z$ boson 
($tZ$) was generated using {\ttf MG5\_aMC@NLO} using the four-flavour PDF scheme. 
Production of a single top quark in the $Wt$-channel together with a \Zboson\ boson ($WtZ$)
was generated with
$\texttt{MG5\_aMC@NLO}$ with the parton shower simulated using
$\texttt{Pythia8}$, the PDF set $\texttt{NNPDF2.3LO}$, and the $\texttt{A14}$ set of tuned parameters.
The diagram removal technique~\cite{DR:Wt} was employed to handle the overlap of $WtZ$ with $t\bar{t}Z$ 
and $t\bar{t}$ production followed by a three-body top-quark decay ($t\to 
WZb$). The procedure also removes the interference between $WtZ$ and these 
two processes. Diboson processes with four charged leptons ($4\ell$), three 
charged leptons and one neutrino ($\ell\ell\ell\nu$), two charged leptons 
and two neutrinos ($\ell\ell\nu\nu$), and diboson processes having 
additional hadronic contributions ($\ell\ell\ell\nu jj$, $\ell\ell\ell\ell 
jj$, $gg\ell\ell\ell\ell$, $\ell\ell\nu\nu jj$) were simulated using the {\ttf 
Sherpa 2.1.1}~\cite{Gleisberg:2008ta} generator. The matrix elements contain all 
diagrams with four electroweak vertices. They were calculated for up 
to one (4$\ell$, $2\ell+2\nu$) or no additional partons ($3\ell+1\nu$) at NLO and up 
to three partons at LO using the {\ttf Comix}~\cite{Gleisberg:2008fv} and 
{\ttf OpenLoops}~\cite{Cascioli:2011va} matrix element generators and were merged 
with the {\ttf Sherpa} parton shower using the {\ttf ME+PS@NLO} prescription~\cite{Cascioli:2011va,Gleisberg:2008fv,Hoeche:2012yf}. 
The {\ttf CT10} PDF set was used in conjunction with a dedicated parton 
shower tuning developed by the {\ttf Sherpa} authors. 
The Higgs boson samples ($\ttbar H$, Higgs boson production via gluon--gluon fusion and
vector boson fusion, and 
in association with a vector boson) were normalized to the theoretical calculations in 
Ref.~\cite{deFlorian:2016spz}.
Events containing $Z$ bosons + jets were simulated with {\ttf 
Powheg-Box v2}~\cite{Alioli:2010xd} interfaced to the {\ttf Pythia8} parton shower model, using 
{\ttf Photos++} version 3.52~\cite{Davidson:2010ew} for QED emissions from 
electroweak vertices and charged leptons. 
The generation of $\ttbar$ and 
single top quarks in the $Wt$-channel was done with {\ttf Powheg-Box v2} and {\ttf Powheg-Box v1}, respectively. 
Due to the high lepton-multiplicity requirement of the event selection
and to increase the sample size, the \ttbar sample was produced by
selecting only true dilepton events in the final state.
The SM production of three or four top quarks and the associated production of a 
\ttbar pair with two $W$ bosons were generated at LO with {\ttf 
MG5\_aMC@NLO+Pythia8}. The production of three massive vector bosons with 
subsequent leptonic decays of all three bosons was modelled at LO with the 
{\ttf Sherpa 2.1.1} generator. Up to two additional partons are included in the 
matrix element at LO.

A set of minimum-bias interactions generated with {\ttf Pythia 8.186} using
the {\ttf A2} set of tuned parameters~\cite{ATL-PHYS-PUB-2012-003} and the
MSTW2008LO~\cite{Martin:2009iq} PDF set were overlaid on the
hard-scattering event to account for additional $pp$ collisions in the
same or nearby bunch crossings (pile-up).
Simulated samples were reweighted to match the pile-up conditions in data.
For most samples, detailed simulation of the detector and trigger system was
performed with standard ATLAS software using {\ttf 
GEANT4}~\cite{Agostinelli:2002hh,SOFT-2010-01}.
Fast simulation based on {\ttf ATLFASTII}~\cite{SOFT-2010-01} is alternatively
used for a few samples dedicated to the evaluation of systematic uncertainties affecting background modelling.
The same offline reconstruction methods used on data are also applied to the samples of simulated events.
Simulated events are corrected so that the object identification,
reconstruction, and trigger efficiencies; the energy scales; and the energy 
resolutions match those determined from data control samples.

\begin{table}
\caption{Generators, parton shower simulation, parton distribution functions, and 
tune parameters used to produce simulated samples for this analysis. 
The acronyms ME and PS stand for matrix element and parton shower, respectively.}
\label{tab:samples}
\begin{center}
\scriptsize
\begin{tabular}{l@{\hskip 6pt}l@{\hskip 6pt}l@{\hskip 6pt}l@{\hskip 6pt}l@{\hskip 6pt}l}
	\hline
	Sample	& Generator & Parton shower & ME PDF & PS PDF & Tune parameters	\\
	\hline
	$\ttbar\to bWqZ$ & {\ttf MG5\_aMC@NLO}~\cite{Alwall:2014hca} & {\ttf Pythia8}~\cite{Sjostrand:2014zea} &
	{\ttf NNPDF3.0NLO}~\cite{Ball:2014uwa} & {\ttf NNPDF2.3LO}~\cite{Ball:2012cx} &  {\ttf A14}~\cite{ATL-PHYS-PUB-2014-021} \\
	$\ttbar V$,$\ttbar H$ & {\ttf MG5\_aMC@NLO} & {\ttf Pythia8} & {\ttf NNPDF3.0NLO} & {\ttf NNPDF2.3LO} &  {\ttf A14} \\
	$\ttbar Z$ (alternative) & {\ttf Sherpa 2.2.0}~\cite{Gleisberg:2008ta} & {\ttf Sherpa 2.2.0} & {\ttf NNPDF3.0NNLO} & {\ttf NNPDF3.0NNLO} &  {\ttf Sherpa} default \\
	$WZ$, $ZZ$ & {\ttf Sherpa 2.1.1} & {\ttf Sherpa 2.1.1} & {\ttf CT10}~\cite{Lai:2010vv} & {\ttf CT10} & {\ttf Sherpa} default \\
	$WZ$ (alternative) & {\ttf Powheg-Box v2}~\cite{Alioli:2010xd} & {\ttf Pythia8} & {\ttf CT10nlo} & {\ttf CTEQ6L1}~\cite{Pumplin:2002vw} & {\ttf AZNLO}~\cite{STDM-2012-23} \\
	$tZ$ & {\ttf MG5\_aMC@NLO} & {\ttf Pythia6}~\cite{Sjostrand:2006za} & {\ttf NNPDF3.0NLO} & {\ttf CTEQ6L1} & {\ttf Perugia2012}~\cite{Skands:2010ak} \\
	$tZ$ (rad. syst.) & {\ttf MG5\_aMC@NLO} & {\ttf Pythia6} & {\ttf NNPDF3.0NLO} & {\ttf CTEQ6L1} & {\ttf P2012RadHi/Lo}~\cite{Skands:2010ak} \\
	$WtZ$ & {\ttf MG5\_aMC@NLO} & {\ttf Pythia8} & {\ttf NNPDF3.0NLO} & {\ttf NNPDF2.3LO} &  {\ttf A14} \\
	$WtZ$ (alternative) & {\ttf MG5\_aMC@NLO} & {\ttf Herwig++}~\cite{Herwig:pp} & {\ttf CT10} & {\ttf CTEQ6L1} &  {\ttf UE-EE-5}~\cite{Herwig:UEEE5} \\
	$gg$F $H$, VBF $H$ & {\ttf Powheg-Box v2} & {\ttf Pythia8} & {\ttf CT10} & {\ttf CTEQ6L1} & {\ttf AZNLO} \\
	$WH$, $ZH$ & {\ttf Pythia8} & {\ttf Pythia8} & {\ttf NNPDF2.3LO} & {\ttf NNPDF2.3LO} & {\ttf A14} \\
	3$t$, 4$t$, $\ttbar WW$  & {\ttf MG5\_aMC@NLO} & {\ttf Pythia8} & {\ttf NNPDF3.0NLO} & {\ttf NNPDF2.3LO} &  {\ttf A14} \\
	Tribosons & {\ttf Sherpa 2.1.1} & {\ttf Sherpa 2.1.1} & {\ttf CT10} & {\ttf CT10} & {\ttf Sherpa} default \\
	$Z+$jets & {\ttf Powheg-Box v2}, & {\ttf Pythia8} & {\ttf CT10} & {\ttf CTEQ6L1} & {\ttf AZNLO} \\
	& {\ttf Photos++}~\cite{Davidson:2010ew} & & & & \\
	$\ttbar\to bWbW$ & {\ttf Powheg-Box v2} & {\ttf Pythia8} & {\ttf CT10} & {\ttf NNPDF2.3LO} & {\ttf A14} \\
	$\ttbar\to bWbW$ (rad. syst.) & {\ttf Powheg-Box v2} & {\ttf Pythia8} & {\ttf CT10} & {\ttf NNPDF2.3LO} & {\ttf A14v3cUp/Do}~\cite{ATL-PHYS-PUB-2014-021} \\
	$\ttbar\to bWbW$ (PS syst.) & {\ttf Powheg-Box v2} & {\ttf Herwig7}~\cite{Herwig:7} & {\ttf NNPDF3.0NLO} & {\ttf MMHT2014lo68cl}~\cite{Harland-Lang2015} & {\ttf H7-UE-MMHT}~\cite{Herwig:7} \\
	$\ttbar\to bWbW$ (ME syst.) & {\ttf MG5\_aMC@NLO} & {\ttf Pythia8} & {\ttf NNPDF3.0NLO} & {\ttf NNPDF2.3LO} &  {\ttf A14} \\
	$Wt$ & {\ttf Powheg-Box v1} & {\ttf Pythia6} & {\ttf CT10f4}  & {\ttf CTEQ6L1} & {\ttf Perugia2012} \\
	\hline
\end{tabular}
\end{center}
\end{table}

%% file: qz13tev-reconstruction.tex
\section{Object reconstruction\label{sec:reco}}

The final states of interest for this search include electrons, muons, 
jets, $b$-tagged jets and missing transverse momentum.

Electron candidates are reconstructed~\cite{PERF-2013-03} from energy 
deposits (clusters) in the electromagnetic calorimeter
matched to reconstructed charged-particle tracks in the inner detector. The 
candidates are required to have a transverse energy $\et>15$~\GeV\ 
and the pseudorapidity of the calorimeter energy cluster associated with the electron 
candidate must satisfy $|\eta_{\mathrm{cluster}}|<2.47$. Clusters in the transition region 
between the barrel and endcap calorimeters, $1.37\leq|\eta_{\mathrm{cluster}}|\leq1.52$,
have poorer energy resolution and are excluded. 
To reduce the background 
from non-prompt sources, electron candidates are also required to satisfy 
$|d_{0}|/\sigma(d_0)<5$ and $|z_0\sin(\theta)|<0.5$~mm criteria, where $d_0$ 
is the transverse impact parameter, with uncertainty 
$\sigma(d_0)$, and $z_0$ is the longitudinal impact parameter
with respect to the primary vertex (defined in Section~\ref{sec:selection}). 
The sum of transverse energies of clusters in the calorimeter 
within a cone of $\Delta R=0.2$ around the electron candidate, 
excluding the $\pT$ of the electron candidate, is required to be less than 6\% of the 
electron $\pT$. The scalar sum of particle transverse momenta around the 
electron candidate within a cone of min(10~\GeV/\pT, 0.2) must be less 
than 6\% of the electron candidate's $\pT$. 

Muon candidates are reconstructed from tracks in the inner detector 
and muon spectrometer, which are combined to improve the reconstruction 
precision and to increase the background rejection~\cite{PERF-2015-10}. They 
are required to have $\pT>15~\GeV$ and $|\eta|<2.5$. Muons are also required to satisfy 
$|d_{0}|/\sigma(d_0)<3$ and $|z_0\sin(\theta)|<0.5$~mm criteria. Additionally, 
the scalar sum of particle transverse momenta around the muon candidate within a 
cone of min(10~\GeV/\pT, 0.3) must be less than 6\% of the muon candidate's 
$\pT$.

Jets are reconstructed from topological clusters of calorimeter cells that 
are noise-suppressed and calibrated to the electromagnetic 
scale~\cite{PERF-2014-07} using the anti-$k_t$ 
algorithm~\cite{Cacciari:2008gp} with a radius parameter $R=0.4$ as 
implemented in FastJet~\cite{Cacciari:2011ma}.
Corrections that change the angles and the energy 
are applied to the jets, starting with a subtraction procedure that uses the jet area to estimate 
and remove the average energy contributed by pile-up interactions~\cite{Cacciari:2007fd}. 
This is followed by a jet-energy-scale 
calibration that restores the jet energy to the mean response of a 
particle-level simulation by correcting variations due to jet flavour and detector geometry
and data driven corrections that match the data to the simulation energy scale~\cite{PERF-2016-04}. Jets 
in the analysis have $\pt>25$~\GeV\ and $|\eta|<2.5$.

To reduce the number of selected jets that originate from pile-up,
an additional selection criterion based on a jet-vertex tagging technique is applied.
Jet-vertex tagging is a likelihood discriminant combining information from several track-based
variables~\cite{PERF-2014-03} and is only applied to
jets with $\pt < 60$~\GeV\ and $|\eta|<2.4$.

Jets containing $b$-hadrons are identified 
($b$-tagged)~\cite{PERF-2012-04,ATL-PHYS-PUB-2016-012} using an algorithm based on 
multivariate techniques. It combines information from the impact parameters 
of displaced tracks and from topological properties of secondary and tertiary 
decay vertices reconstructed within the jet. 
Using simulated  \ttbar\ events, the $b$-tagging efficiency for jets originating from a $b$-quark is 
determined to be 77\% for the chosen working point, while the rejection factors for 
light-flavour jets and charm jets are 134 and 6, respectively.

The missing transverse momentum $\vec{p}^\textrm{miss}_\textrm{T}$ is
the negative vector sum of the $\pT$ of all selected and 
calibrated objects in the event, including a term to account for
soft particles in the event that are not associated with any of the selected 
objects~\cite{PERF-2011-07,PERF-2014-04}. 
To reduce contamination from pile-up interactions, the soft term is calculated from 
inner detector tracks matched to the selected primary vertex.
The magnitude of the missing transverse momentum is \MET.

To avoid double counting of single final-state objects, such as an 
isolated electron being reconstructed as both an electron and a jet,
the following procedures are applied in the order given.
Electron candidates which share a track with a 
muon candidate are removed. If the distance in $\Delta R$ between a jet and an 
electron candidate is $\Delta R < 0.2$, then the jet is dropped. 
If, for the same electron, multiple
jets are found with this requirement, only the closest one is dropped. If the 
distance in $\Delta R$ between a jet and an electron is $0.2 < \Delta 
R < 0.4$, then the electron is dropped. If the distance in $\Delta R$ between 
a jet and a muon candidate is $\Delta R < 0.4$ and if the jet has more than 
two associated tracks, the muon is dropped; otherwise the jet is removed.

%% file: qz13tev-selection.tex
\section{Event selection and reconstruction\label{sec:selection}}

Events considered in this analysis meet the following criteria.
At least one of the selected leptons must be matched, with $\Delta R < 0.15$, 
to the appropriate trigger object and have transverse momentum greater than 
25~\GeV\ or 27~\GeV\ for the data collected in 2015 or 2016, respectively.
The events must have at least one primary vertex.
The primary vertex must have at least two associated tracks, each with 
$\pt>400$~MeV. The primary vertex 
with the highest sum of $\pt^2$ over all associated tracks is chosen.
Exactly three isolated charged leptons with $|\eta|<2.5$ and $\pt > 15$~\GeV\ are 
required. The $Z$ boson candidate is reconstructed from the two leptons that 
have the same flavour, opposite charge, and a reconstructed mass within 
15~\GeV\ of the $Z$ boson mass ($m_Z$).
If more than one compatible lepton pair is found, the one with the reconstructed mass closest 
to $m_Z$ is chosen as the $Z$ boson candidate. According to the signal 
topology, the events are then required to have $\MET>20$~\GeV\ and at least 
two jets. All jets must have $\pt > 25$~\GeV\ and $|\eta|<2.5$, and 
exactly one of the jets must be $b$-tagged.

Applying energy--momentum conservation, the kinematic properties of the top quarks 
are reconstructed from the corresponding decay particles by minimizing
\begin{equation*}
\chi^2  =  \frac{\left(m^{\mathrm{reco}}_{j_a\ell_a\ell_b}-m_{t_{\mathrm{FCNC}}}\right)^2}{\sigma_{t_{\mathrm{FCNC}}}^2}+
\frac{\left(m^{\mathrm{reco}}_{j_b\ell_c\nu}-m_{t_{\mathrm{SM}}}\right)^2}{\sigma_{t_{\mathrm{SM}}}^2}
 + \frac{\left(m^{\mathrm{reco}}_{\ell_c\nu}-m_W\right)^2}{\sigma_W^2},
\label{eq:chi2}
\end{equation*}
where $m^{\mathrm{reco}}_{j_a\ell_a\ell_b}$, 
$m^{\mathrm{reco}}_{j_b\ell_c\nu}$, and $m^{\mathrm{reco}}_{\ell_c\nu}$ are 
the reconstructed masses of the $qZ$, $bW$, and $\ell\nu$ systems, respectively. 
For each jet combination, $j_b$ corresponds to the $b$-tagged jet, while any jet can be assigned to $j_a$. 
Since the neutrino from the semileptonic decay of the top quark ($t\to bW\to b\ell\nu$) is 
undetected, its four-momentum must be estimated. This is done by 
assuming that the lepton not previously assigned to the $Z$ boson and the 
$b$-tagged jet originate from the $W$ boson and SM top-quark decay,
respectively, and that $\vec{p}^\textrm{miss}_\textrm{T}$ is the transverse momentum vector of the neutrino in the $W$ boson decay.
The longitudinal component of the neutrino momentum ($p^{\nu}_z$) is then determined
by the minimization of the $\chi^2$.
The central values of the masses and the widths of the 
top quarks and the $W$ boson are taken from simulated signal 
events. This is done by matching the particles in the simulated events  
to the reconstructed ones, setting the longitudinal momentum of the 
neutrino to the $p_z$ of the simulated neutrino, and then performing 
Bukin fits\footnote{These fits use a piecewise function with  a Gaussian function in 
the centre and two asymmetric tails. Six parameters determine the overall normalization, the peak position, the
width of the core, the asymmetry, the size of the lower tail, and the size of the higher tail.
Of these, only the peak position and the width enter the $\chi^2$.}~\cite{Bukin} to the masses of the 
matched reconstructed top quarks and $W$ boson. The obtained values are 
$m_{t_{\mathrm{FCNC}}}=169.6$~\GeV, $\sigma_{t_{\mathrm{FCNC}}}=12.0$~\GeV, 
$m_{t_{\mathrm{SM}}}=167.2$~\GeV, $\sigma_{t_{\mathrm{SM}}}=24.0$~\GeV, 
$m_W=81.2$~\GeV\ and $\sigma_W=15.1$~\GeV.
The $\chi^2$ minimization gives the most
probable value for $p^{\nu}_z$ for a given combination. 
The combination with the minimum $\chi^2$ is chosen, 
which fixes the assignment of reconstructed particles and the corresponding $p^{\nu}_z$ value.
The jet from the top-quark FCNC decay is referred to as the light-quark ($q$) 
jet. The fractions of correct assignments between the reconstructed top 
quarks and the true simulated particles (evaluated as a match within a cone 
of size $\Delta R = 0.4$) are $\epsilon_{t_{\mathrm{FCNC}}}=80$\% and 
$\epsilon_{t_{\mathrm{SM}}}=58$\%, where the difference comes from the fact that for the SM top-quark decay 
the match of the $\MET$ with the simulated neutrino is less efficient. 

\begin{figure}
\begin{center}

  \includegraphics[width=.48\textwidth]{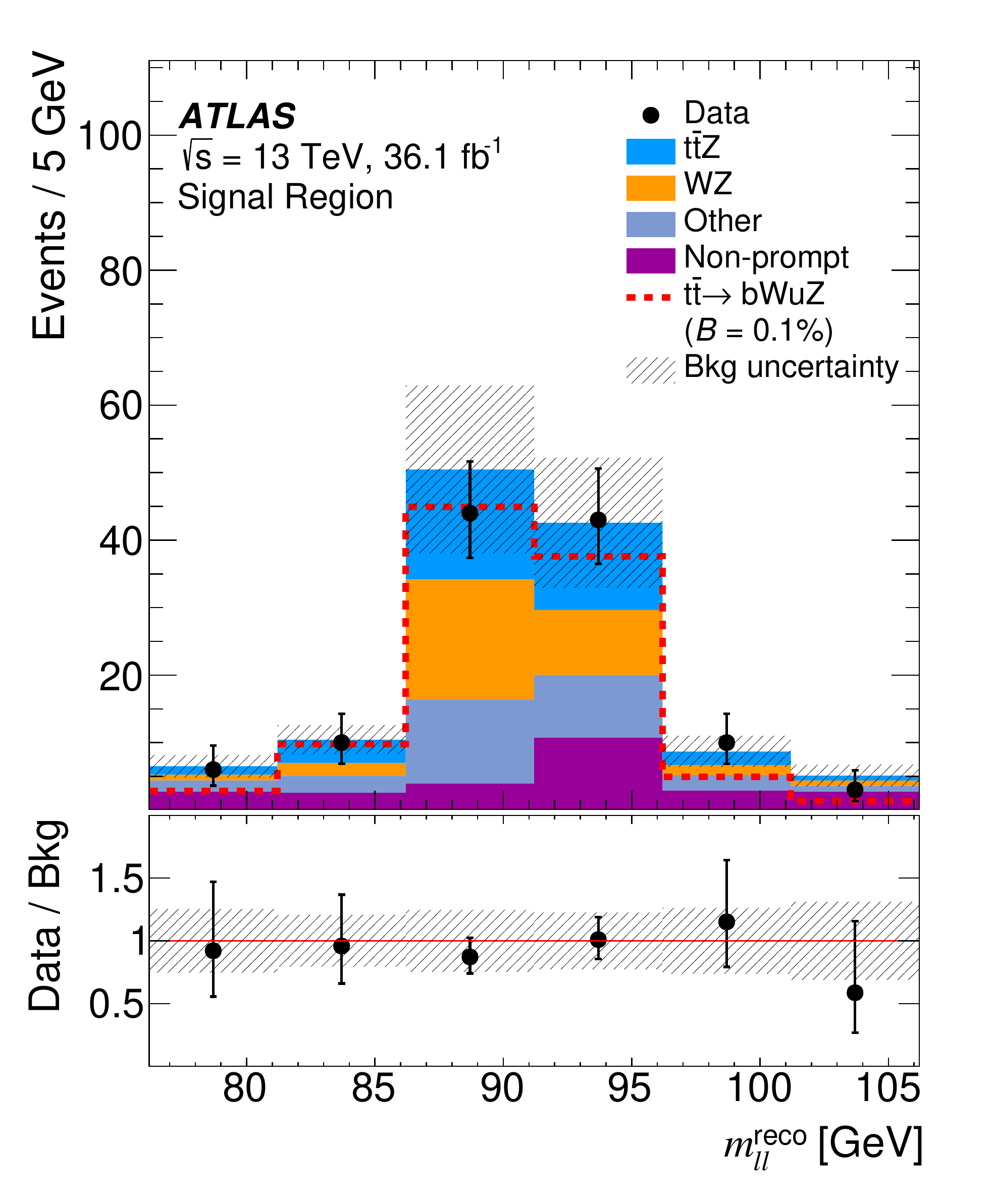}
  \includegraphics[width=.48\textwidth]{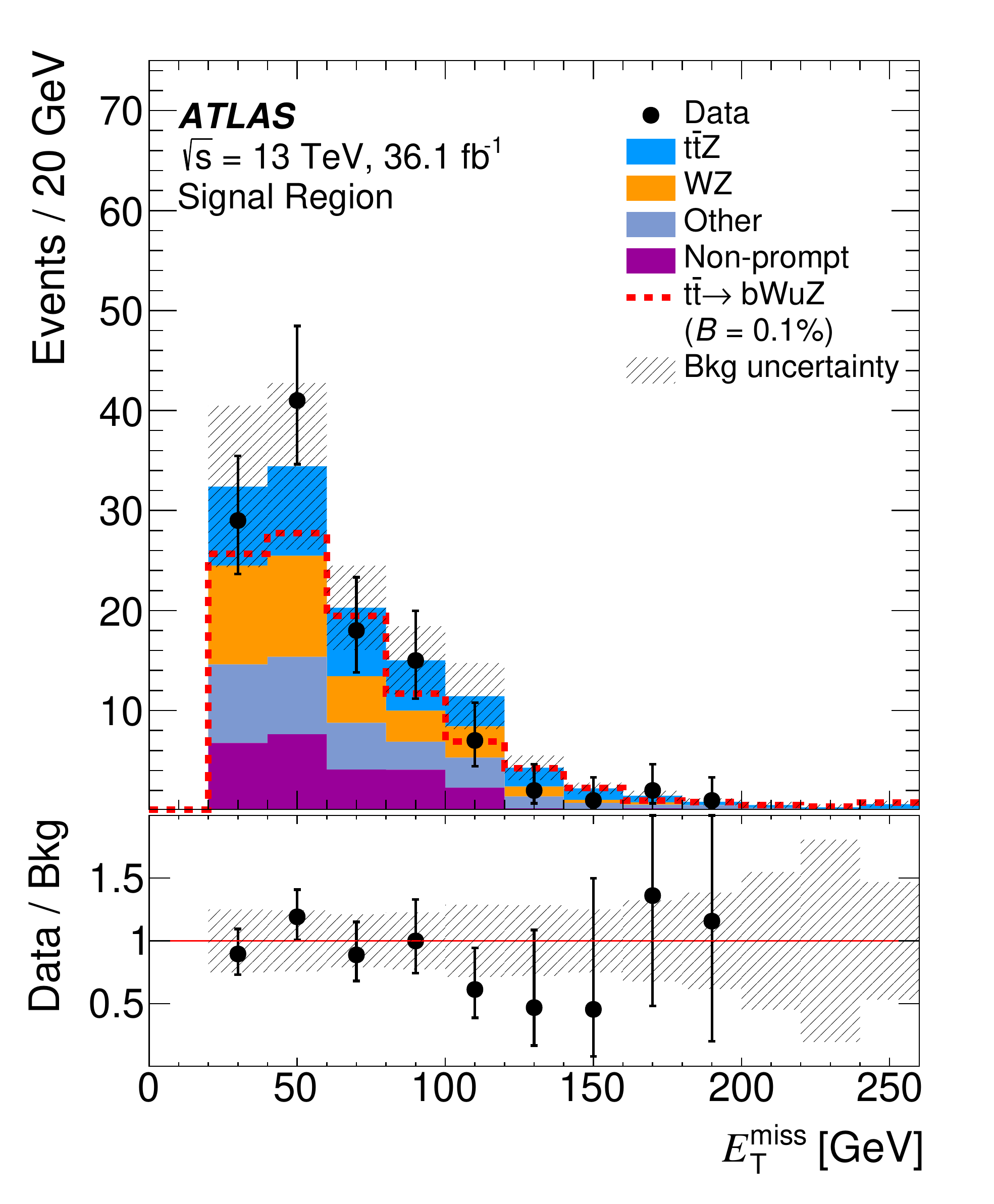}

  (a)\hspace*{.48\textwidth}(b)

  \includegraphics[width=.48\textwidth]{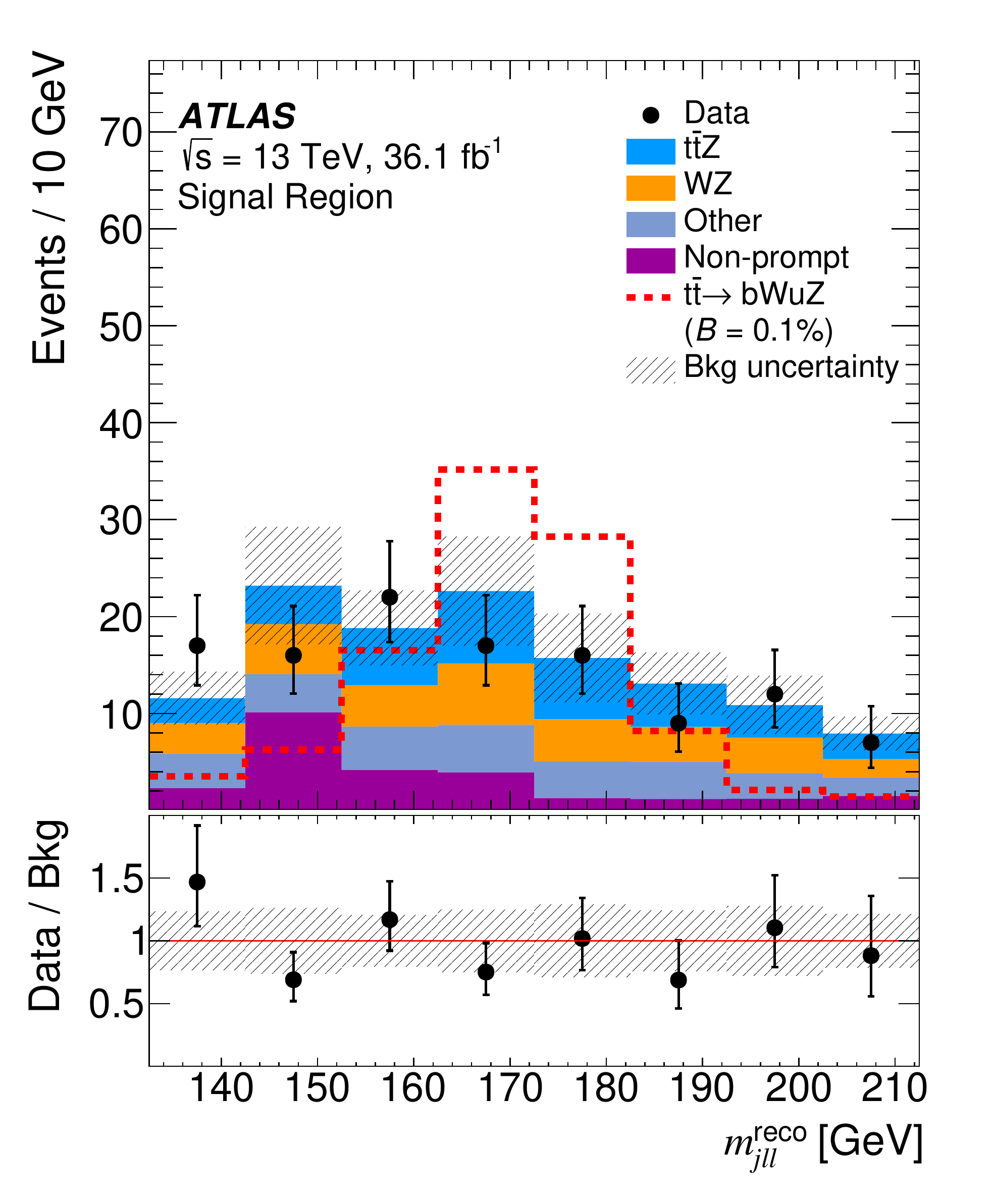}
  \includegraphics[width=.48\textwidth]{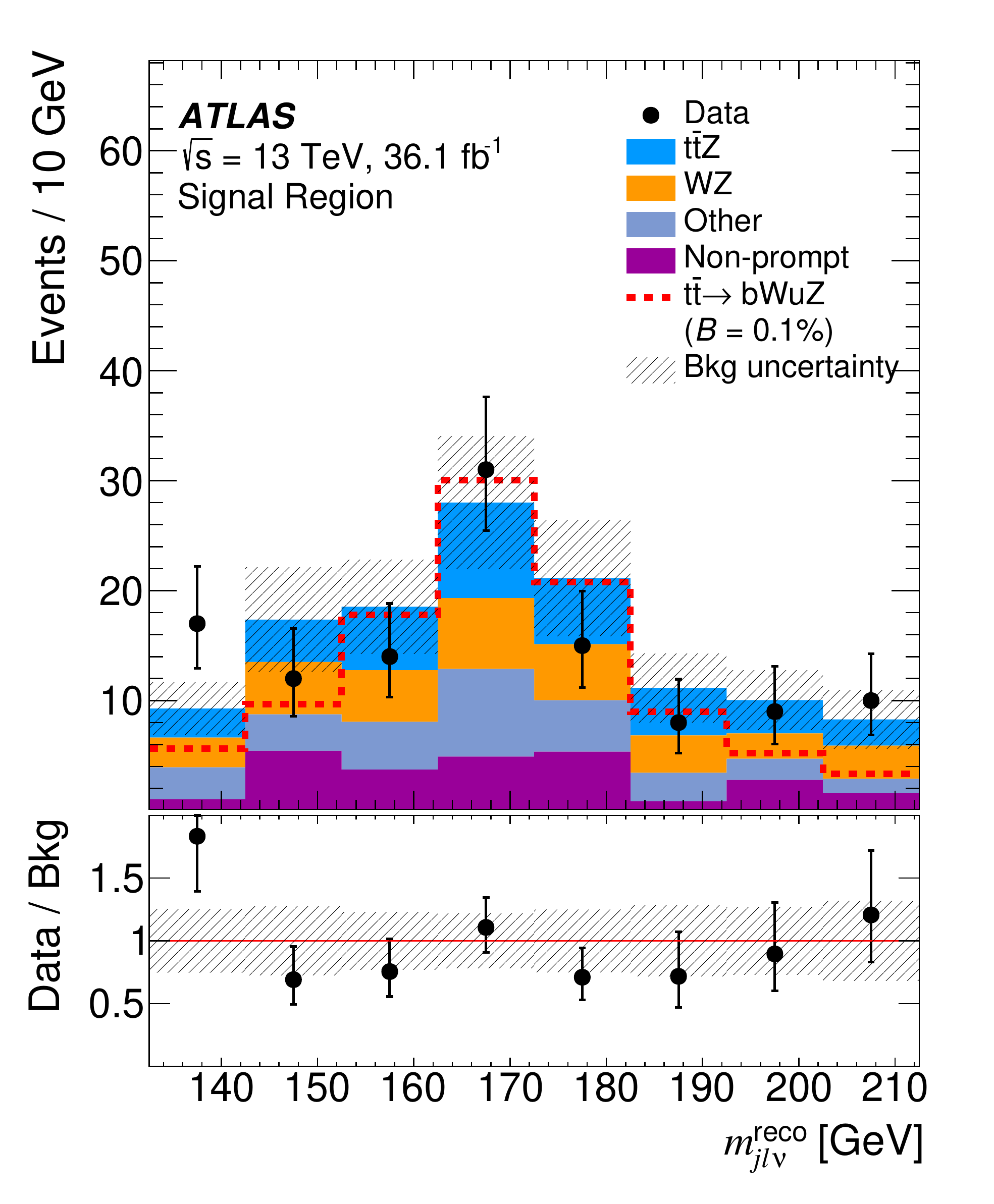}

  (c)\hspace*{.48\textwidth}(d)
\end{center}
  \caption{Expected (filled histogram) and observed (points with error bars) 
distributions in the SR before the combined fit under the background-only hypothesis of 
  (a) the mass of the $Z$ boson candidate,
  (b) \MET,
  (c) the mass of the top-quark candidate with FCNC decay, and
  (d) the mass of the top-quark candidate with SM decay.
  For comparison, distributions for the FCNC $\ttbar \to bWuZ$ signal (dashed 
  line), normalized to ${\mathcal{B}}$(\tuz) = 0.1\%, are also shown.
  The dashed area represents the total uncertainty in the background prediction.
  The first (last) bin in all distributions includes the underflow (overflow).
  The ``Other'' category includes all remaining backgrounds described in 
  Section~\ref{sec:samples}. 
}
  \label{fig:preselection}
\end{figure}

The final requirements defining the signal region (SR) are 
|$m^{\mathrm{reco}}_{j_a\ell_a\ell_b} - 172.5~\GeV|<40$~\GeV, 
|$m^{\mathrm{reco}}_{j_b\ell_c\nu} - 172.5~\GeV|<40$~\GeV, and 
$|m^{\mathrm{reco}}_{\ell_c\nu} - 80.4~\GeV|<30$~\GeV. 
Figure~\ref{fig:preselection} shows the mass of the $Z$ boson candidate as 
well as the \MET\ and the masses of both top-quark candidates for the events 
fulfilling these requirements. The stacked histograms show backgrounds with three prompt
leptons, normalized to the theory predictions, and the scaled background from 
non-prompt leptons, normalized as discussed in the next section.

%% file: qz13tev-backgrounds.tex
\section{Background estimation and control regions\label{sec:backgrounds}}

The main sources of background events containing three prompt leptons 
are: diboson production, \ttz, and $tZ$ processes. 
In addition, events where one or more of the reconstructed leptons are non-prompt, either
mis-reconstructed or from heavy-flavour decays, are
background sources.
To assess how well the data agree with the simulated samples of the 
expected background, five control regions~(CRs) are defined and 
included in the final fit.  This allows
rescaling of the background expectations to the best fit with
observed data and reduces the background
uncertainties.  Systematic uncertainties in the signal yield are also
reduced by the final fit (Section~\ref{sec:limits}).

Backgrounds from events containing at least one non-prompt lepton are 
estimated by means of a semi-data-driven technique using dedicated selections.
This technique uses the data to determine the
normalization for simulated $Z$+jets and \ttbar events with a non-prompt electron and non-prompt muon separately.
In order to determine the non-prompt lepton scale factors ($\lambda^{e}$, $\lambda^{\mu}$) for simulated $Z$+jets and \ttbar events,
four regions are defined each enriched with non-prompt electrons
or muons from $Z$+jets events (``light'' region) or \ttbar events (``heavy'' region).
The selections used, which are given in Table~\ref{tab:backgrounds:FakeCRs}, make the four regions orthogonal to the  CRs and SR used in the final fit (Section~\ref{sec:limits}).
The non-prompt lepton scale factors for $Z$+jets and \ttbar events 
are expected to be different due to differences in background sources and are 
determined by a simultaneous 
likelihood fit to the inclusive yields in the four regions, taking into account statistical 
and systematic uncertainties, leading to 
$\lambda_{Z+\mathrm{jets}}^{e}=2.2 \pm 0.8$, $\lambda_{Z+\mathrm{jets}}^{\mu} = 1.9 \pm 0.9$, $\lambda_{\ttbar}^{e} = 
1.1 \pm 0.3$, and $\lambda_{\ttbar}^{\mu} = 1.1 \pm 0.7$.
These non-prompt lepton scale factors are applied to the $Z$+jets and \ttbar samples
in the CRs and SR used in the final fit (Section~\ref{sec:limits}).
Agreement between data and 
expectations in the CRs is significantly improved after applying the non-prompt lepton scale 
factors to the simulated samples.

\begin{table}[!bp]
    \caption{Selection criteria applied to derive the four scale factors
    of the non-prompt leptons background.
    OS indicates a pair of opposite-sign leptons, OSSF 
    indicates a pair of opposite-sign, same-flavour leptons. 
    Additionally, events with at least two jets, one of them 
    $b$-tagged, and 20~\GeV\ < $\MET$ < 40~\GeV\ in the SR
    are rejected from the ``light'' regions.}
    \label{tab:backgrounds:FakeCRs}
	\begin{center}
		\begin{tabular}{cccc}
			\hline
			``Light'' region -- $e$ &  ``Light'' 
			region -- $\mu$ &
			``Heavy'' region -- $e$ & ``Heavy'' 
			region -- $\mu$ \\
			\hline
			$eee$ or $e\mu \mu$, OSSF &
			$\mu\mu\mu$ or $\mu ee$, OSSF &
			$e\mu \mu$ , OS no OSSF&
			$\mu ee$, OS no OSSF \\
			\small{$|m_{\ell \ell} - 91.2~\text{GeV}| < 15~\text{GeV}$} &
			\small{$|m_{\ell \ell} - 91.2~\text{GeV}| < 15~\text{GeV}$} & &\\
			{$\geq$ 1 jet}& {$\geq$ 1 jet} &  $\geq$ 2 jet &
			$\geq$ 2 jet \\
			{$\MET < 40$~\GeV } &{$\MET < 40$~\GeV } && \\			
			{$ m_{\mathrm{T}^{\phantom{2}}} \leq 50$~\GeV} &
			{$ m_{\mathrm{T}^{\phantom{2}}} \leq 50$~\GeV }& & \\
			\hline
		\end{tabular}
	\end{center}
\end{table}

The \ttz\ CR requires exactly three leptons, two 
of them with the same flavour, opposite charge, and a reconstructed $m_{\ell \ell}$ within 15~\GeV\ of the $Z$ boson mass. Furthermore, the events are required to 
have at least four jets with $\pt > 25$~\GeV\ and $|\eta|<2.5$, two of which 
must be $b$-tagged.

The $WZ$ CR requires three leptons, two of them with the same flavour, 
opposite charge, and a reconstructed $m_{\ell \ell}$ within 15~\GeV\ of the $Z$ boson 
mass. Additional requirements are the presence of at least two jets 
with $\pt > 25$~\GeV\ and $|\eta|<2.5$, the leading jet 
having $\pt > 35$~\GeV, no $b$-tagged jets with $\pt > 25$~\GeV, $\MET 
>40$~\GeV, and a transverse mass 
$m_{\text{T}}^{\ell \nu} > 50$~\GeV, where $m_{\text{T}}^{\ell \nu}$
is calculated from the momentum of the non-\Zboson\ lepton and the missing
transverse momentum vector.

The $ZZ$ CR requires two pairs of leptons, each with the 
same flavour, opposite charge, and a reconstructed $m_{\ell \ell}$ within 15~\GeV\ of the 
$Z$ boson mass. At least one jet with $\pt > 25$~\GeV\ and $|\eta|<2.5$, no 
$b$-tagged jets with $\pt > 25$~\GeV, and $\MET >20$~\GeV\ are also required.

The CR for the non-prompt 
lepton backgrounds requires three leptons with two of them having the same 
flavour, opposite charge, and a reconstructed $m_{\ell \ell}$ outside $15~\GeV$ of the $Z$ 
boson mass, at least one jet with $\pt > 25$~\GeV, and $\MET > 20~\GeV$. This CR is split into 
two regions, with either zero (CR0) or exactly one (CR1) $b$-tagged jet. 

Table~\ref{tab:backgrounds:requirements} summarizes the selection 
requirements described above. The expected and observed yields in these 
regions, before the background fit described in Section~\ref{sec:limits}, are shown in 
Table~\ref{tab:backgrounds:beforefit}.

\begin{table}
\caption{The selection requirements applied for the background control and signal regions.
OSSF refers to the presence of a pair of opposite-sign, same-flavour leptons.}
\label{tab:backgrounds:requirements}    
\begin{center}
\begin{tabular}{lccccc}
\hline
Selection  & $\ttbar Z$ CR & $WZ$ CR & $ZZ$ CR & Non-prompt lepton CR0 (CR1)& SR \\
\hline
No. leptons  & 3 & 3 & 4 & 3 & 3 \\
OSSF & Yes & Yes & Yes & Yes & Yes \\
$|m^{\mathrm{reco}}_{\ell \ell} - 91.2~\mathrm{GeV}|$ &  $< 15$~\GeV &  $< 15$~\GeV &  $< 15$~\GeV &  $> 15$~\GeV &  $< 15$~\GeV \\
No. jets & $\geq 4$ & $\geq 2$ & $\geq 1$ & $\geq 2$ & $\geq 2$\\
No. $b$-tagged jets & 2 & 0 & 0 & 0 (1) & 1 \\
$\MET$ & $> 20$~\GeV & $> 40$~\GeV & $> 20$~\GeV & $> 20$~\GeV & $> 20$~\GeV \\
$m_{\text{T}}^{\ell \nu}$ & - & $> 50$~\GeV & - & - & - \\
$|m_{\ell \nu}^{\text{reco}} - 80.4~\mathrm{GeV}|$ & - & - & - & - & $< 30$~\GeV \\
$|m_{j \ell \nu}^{\text{reco}} - 172.5~\mathrm{GeV}|$ & - & - & - & - & $< 40$~\GeV \\
$|m_{j \ell \ell}^{\text{reco}} - 172.5~\mathrm{GeV}|$ & - & - & - & - & $< 40$~\GeV \\
\hline
\end{tabular}
\end{center}    
\end{table}

\begin{table}
\caption{Event yields in the background CRs for all significant 
sources of events before the combined fit under the background-only hypothesis
described in Section~\ref{sec:limits}.
The uncertainties shown include all of the systematic uncertainties described in Section~\ref{sec:systematics}. 
The entry labelled ``other backgrounds'' includes all remaining backgrounds described in Section~\ref{sec:samples} 
and in Table~\ref{tab:samples}.}
\label{tab:backgrounds:beforefit}
\begin{center}
	\begin{tabular}{lr@{\,}@{$\pm$}@{\,}lr@{\,}@{$\pm$}@{\,}lr@{\,}@{$\pm$}@{\,}lr@{\,}@{$\pm$}@{\,}lr@{\,}@{$\pm$}@{\,}l}
		\hline
		Sample  	& \multicolumn{2}{c}{\ttz\ CR} & \multicolumn{2}{c}{$WZ$ CR} & \multicolumn{2}{c}{$ZZ$ CR}	& \multicolumn{2}{c}{Non-prompt} 		& \multicolumn{2}{c}{Non-prompt}	\\
		&\multicolumn{2}{c}{}&\multicolumn{2}{c}{}&\multicolumn{2}{c}{}&  \multicolumn{2}{c}{lepton CR0} & \multicolumn{2}{c}{lepton CR1} \\
		\hline		
		\ttz & \num[round-mode=figures,round-precision=2]{60.8441} & \num[round-mode=figures,round-precision=1]{8.71178} & \num[round-mode=figures,round-precision=3]{16.3393} & \num[round-mode=figures,round-precision=2]{3.1297} & \num[round-mode=figures,round-precision=2]{0} & \num[round-mode=figures,round-precision=2]{0} & \num[round-mode=figures,round-precision=2]{6.07682} & \num[round-mode=figures,round-precision=2]{1.23666} & \num[round-mode=figures,round-precision=3]{22.0864} & \num[round-mode=figures,round-precision=2]{3.18934}  \\
		$WZ$ & \num[round-mode=figures,round-precision=1]{8.94761} & \num[round-mode=figures,round-precision=1]{8.74568} & \num[round-mode=figures,round-precision=2]{559.373} & \num[round-mode=figures,round-precision=2]{237.809} & \num[round-mode=figures,round-precision=2]{0} & \num[round-mode=figures,round-precision=2]{0} & \num[round-mode=figures,round-precision=2]{150.275} & \num[round-mode=figures,round-precision=1]{67.8302} & \num[round-mode=figures,round-precision=2]{20.3514} & \num[round-mode=figures,round-precision=1]{9.24465}  \\
		$ZZ$ & \num[round-mode=figures,round-precision=1]{0.0703854} & \num[round-mode=figures,round-precision=1]{0.0264837} & \num[round-mode=figures,round-precision=2]{48.315} & \num[round-mode=figures,round-precision=2]{10.6594} & \num[round-mode=figures,round-precision=2]{91.8011} & \num[round-mode=figures,round-precision=2]{20.4229} & \num[round-mode=figures,round-precision=2]{58.0254} & \num[round-mode=figures,round-precision=2]{15.6408} & \num[round-mode=figures,round-precision=2]{9.01507} & \num[round-mode=figures,round-precision=2]{2.28046}  \\
		Non-prompt leptons & \num[round-mode=figures,round-precision=1]{3.28429} & \num[round-mode=figures,round-precision=1]{5.53569} & \num[round-mode=figures,round-precision=2]{28.4379} & \num[round-mode=figures,round-precision=2]{16.0779} & \num[round-mode=figures,round-precision=2]{0} & \num[round-mode=figures,round-precision=2]{0} & \num[round-mode=figures,round-precision=2]{149.941} & \num[round-mode=figures,round-precision=1]{50.33} & \num[round-mode=figures,round-precision=2]{142.946} & \num[round-mode=figures,round-precision=1]{68.5264}  \\
		Other backgrounds & \num[round-mode=figures,round-precision=3]{13.3989} & \num[round-mode=figures,round-precision=2]{2.72026} & \num[round-mode=figures,round-precision=2]{21.8027} & \num[round-mode=figures,round-precision=1]{5.35549} & \num[round-mode=figures,round-precision=1]{0.96733} & \num[round-mode=figures,round-precision=1]{0.602068} & \num[round-mode=figures,round-precision=2]{17.4018} & \num[round-mode=figures,round-precision=1]{6.17267} & \num[round-mode=figures,round-precision=2]{32.0885} & \num[round-mode=figures,round-precision=1]{6.48363} \\
		\hline
		Total background & \num[round-mode=figures,round-precision=2]{86.5453} & \num[round-mode=figures,round-precision=2]{14.8693} & \num[round-mode=figures,round-precision=2]{674.268} & \num[round-mode=figures,round-precision=2]{241.317} & \num[round-mode=figures,round-precision=2]{92.7685} & \num[round-mode=figures,round-precision=2]{20.4332} & \num[round-mode=figures,round-precision=2]{381.72} & \num[round-mode=figures,round-precision=1]{91.6786} & \num[round-mode=figures,round-precision=2]{226.487} & \num[round-mode=figures,round-precision=1]{69.7111} \\
		\hline 
		Data &	\multicolumn{2}{c}{81}	& \multicolumn{2}{c}{734}	& \multicolumn{2}{c}{87}	& \multicolumn{2}{c}{433\pho}	& \multicolumn{2}{c}{260\pho} \\
		\hline
		Data / Bkg   & \num[round-mode=figures,round-precision=2]{0.935926} & \num[round-mode=figures,round-precision=2]{0.191498} & 
		\num[round-mode=figures,round-precision=2]{1.08859} & \num[round-mode=figures,round-precision=1]{0.391666} & \num[round-mode=figures,round-precision=2]{0.937819} & \num[round-mode=figures,round-precision=2]{0.229735} & 
		\num[round-mode=figures,round-precision=3]{1.13434} & \num[round-mode=figures,round-precision=2]{0.277837} &
		\num[round-mode=figures,round-precision=2]{1.14797} & \num[round-mode=figures,round-precision=1]{0.360437} \\ 
		\hline
	\end{tabular}
\end{center}
\end{table}

%% file: qz13tev-systematics.tex
\section{Systematic uncertainties\label{sec:systematics}}

The background fit to the CRs, described in 
Section~\ref{sec:limits}, reduces 
the systematic uncertainty from some sources, due to the constraints introduced by the data.
The effect on shape and normalization of each source of systematic uncertainty before the fit is studied by 
independently varying each parameter within its estimated uncertainty
and propagating this through the full analysis chain. 

The main uncertainties, in both the background and signal estimations, are
from theoretical normalization uncertainties
and uncertainties in the modelling of background processes in the simulation.

The theoretical normalization uncertainties are estimated to be 12\% for
\ttz, 13\% for $t\bar{t}W$~\cite{Alwall:2014hca}, and 30\%
for $tZ$ production~\cite{CERN-EP-2017-188}.
For dibosons, the uncertainties in the normalization
of the cross section~\cite{STDM-2015-19} and from the choice of values for the
electroweak parameters~\cite{Hoche:2010kg}
are added in quadrature, yielding a 12.5\% uncertainty.
An uncertainty of $+10$\% and $-28$\% is assigned to the $WtZ$ background
cross section following the methodology of 
Ref.~\cite{TOPQ-2015-22}.
For the remaining small backgrounds, a 50\% uncertainty is assumed.
The $\ttbar$ production cross-section uncertainties from
the independent variation of the factorization and renormalization scales,
the PDF choice, and $\alpha_\mathrm{S}$ variations
(see Refs.~\cite{Czakon:2011xx,Botje:2011sn} and references 
therein and Refs.~\cite{Martin:2009bu,Gao:2013xoa,Ball:2012cx})
give a 5\% uncertainty in the signal normalization.

The uncertainties in the modelling of \ttz\ and $WZ$ processes in the simulation
are taken from alternative
generators ({\ttf Sherpa 2.2.0} and {\ttf Powheg-Box v2}  interfaced to
the {\ttf Pythia8}, respectively) which yield 4\% and 50\% uncertainties in the SR, respectively.
The $WtZ$ parton-shower uncertainty is estimated as 6\% in the SR by using a sample interfaced to {\ttf Herwig++}~\cite{Herwig:pp}. 
The effect of QCD radiation on the $tZ$ production process is estimated to be below 2\% in the SR by using  alternative {\ttf MG5\_aMC@NLO\_Pythia6}~\cite{Sjostrand:2006za} $tZ$ samples with additional radiation.
The uncertainty due to the choice of NLO generator for the \ttbar\
event production is evaluated using the alternative sample generated
with {\ttf MG5\_aMC@NLO} interfaced to {\ttf Pythia8}.
The uncertainty in the total non-prompt leptons background in the SR is 25\%.
To evaluate the uncertainty due to the choice of parton shower
algorithm, ~$\ttbar$ samples generated using {\ttf Powheg} interfaced
to {\ttf Herwig7}~\cite{Herwig:7} are used, yielding 2\% uncertainty in the total non-prompt leptons
background in the SR. To estimate the effect of QCD
radiation on the $\ttbar$ samples, alternative samples generated with
{\ttf Powheg+Pythia8} are considered where the factorization and
renormalization scales are varied up and down by a factor of two 
and the A14 set of tuned parameters is changed to a version that varied the {\ttf VAR3c}~\cite{ATL-PHYS-PUB-2014-021}
parameter, changing the amount of QCD radiation.
This leads to a 10\% uncertainty in the total non-prompt leptons background in the SR.
Non-prompt lepton scale factor uncertainties are considered in the estimation of the backgrounds from events
containing at least one non-prompt lepton.

For both the estimated signal and background event yields, experimental 
uncertainties resulting from detector effects are considered, including the lepton reconstruction, identification and trigger efficiencies, as well as lepton momentum scales and resolutions~\cite{PERF-2013-03, PERF-2013-05, PERF-2014-05}. Uncertainties of the $\MET$ scale~\cite{PERF-2011-07}, pile-up effects, and jet-energy scale and 
resolution~\cite{PERF-2012-01, PERF-2011-04} are also considered.  The $b$-tagging 
uncertainty component, which includes the uncertainty of the $b$-, $c$-, 
mistagged- and $\tau$-jet scale factors (the $\tau$ and charm uncertainties 
are highly correlated and evaluated as such) is evaluated by varying the 
$\eta$-, \pt- and flavour-dependent scale factors applied to each jet in the 
simulated samples. 
The relative impact of each type of systematic 
uncertainty on the total background and signal yields is summarized
before and after the fit in Table~\ref{tab:fit:systs_preFit} and Table~\ref{tab:fit:systs_postFit}, respectively.

The uncertainty related to the integrated luminosity for the dataset used in 
this analysis is 2.1\%. It is derived following the methodology described in 
Ref.~\cite{DAPR-2010-01} and only affects background estimates from 
simulated samples.

\begin{table}[htbp]
	\caption{Summary of the relative impact of each type of uncertainty on the total background (B) yield in the background control regions and on the background and signal (S) yields in the signal region before the combined fit under the background-only hypothesis. }
	\label{tab:fit:systs_preFit} 
	\begin{center}
		\small
		\begin{tabular}{lS[table-format=3.2]S[table-format=3.2]S[table-format=3.2]S[table-format=3.2]S[table-format=3.2]S[table-format=3.2]S[table-format=3.2]}
			\hline 
			Pre-fit & {$t\bar{t}Z$ CR} & {$WZ$ CR} & {$ZZ$ CR}  & 
			{Non-prompt} & {Non-prompt} & \multicolumn{2}{c}{SR}  \\
			Source & & & & {lepton CR0} & {lepton CR1} && \\
			 & {B [\%]} & {B [\%]} & {B [\%]} & {B [\%]} & {B [\%]} & {B [\%]} & {S [\%]}\\
			\hline 
			Event modelling & 29 &	40 &	13 &	24 &	40 &	30 &	5 \\
			Leptons & 2.1 &	2.4 &	3.0 &	2.6 &	2.9 &	2.6 &	1.9 \\
			Jets & 6 &	8 &	15 &	10 &	4 &	9 &	4 \\
			$b$-tagging &	 7	& 1.5 &	0.6 &	2.3 &	3.0 &	5 &	3.4 \\
			$\MET$ & 0.4 &	4 &	2.6 &	3.0 &	0.8 &	5 &	1.4 \\
			Non-prompt leptons &	1.1 &	1.3 &	{---} &	12 &	15 &	6 &	{---} \\
			Pile-up & 5	& 1.3 &	5 &	3.5 &	1.8 &	4 &	2.3  \\
			Luminosity & 2.0 &	2.0 &	2.1 &	1.3 &	0.8 &	1.7 &	2.1 \\
			\hline 
		\end{tabular} 
	\end{center} 
\end{table} 

\begin{table}[htbp]
	\caption{Summary of the relative impact of each type of uncertainty on the total background (B) yield in the background control regions and on the background and signal (S) yields in the signal region after the combined fit under the background-only hypothesis.}
	\label{tab:fit:systs_postFit} 
	\begin{center}
		\small
		\begin{tabular}{lS[table-format=3.2]S[table-format=3.2]S[table-format=3.2]S[table-format=3.2]S[table-format=3.2]S[table-format=3.2]S[table-format=3.2]}
			\hline 
			Post-fit & {$t\bar{t}Z$ CR} & {$WZ$ CR} & {$ZZ$ CR}  & 
			{Non-prompt} & {Non-prompt} & \multicolumn{2}{c}{SR}  \\
			Source & & & & {lepton CR0} & {lepton CR1} && \\
			& {B [\%]} & {B [\%]} & {B [\%]} & {B [\%]} & {B [\%]} & {B [\%]} & {S [\%]}\\
			\hline
			Event modelling & 22 &	10 &	11 &	9 &	23 &	18 &	5 \\
			Leptons & 2.0 &	2.4 &	2.9 &	2.6 &	2.9 &	2.6 &	1.8 \\
			Jets & 5 &	6 &	11 &	8 &	4 &	8 &	4 \\
			$b$-tagging &	7 &	1.4 &	0.6 &	2.1 &	2.8 &	4 &	3.1 \\
			$\MET$ & 0.3 &	3.3 &	2.5 &	2.8 &	0.7 &	4 &	1.4 \\
			Non-prompt leptons 	&1.1 &	1.1 &  {---} &	8 &	12 &	5 &	{---} \\
			Pile-up & 	5 &	1.2 &	5 &	3.3 &	1.7 &	3.5 &	2.2 \\
			Luminosity & 2.0 &	2.0 &	2.1 &	1.3 &	0.8 &	1.6 &	2.1 \\
			\hline  
		\end{tabular} 
	\end{center} 
\end{table} 

%% file: qz13tev-limits.tex
\section{Results\label{sec:limits}}

A simultaneous fit to the SR and all CRs defined in
Table~\ref{tab:backgrounds:requirements} is used to search for a
signal from FCNC decays of the top quark.
A maximum-likelihood fit is performed to kinematic distributions in
the signal and control regions to test for the presence of signal
events.
Contamination of the CRs by the signal is negligible.
The inclusion of the CRs in a combined fit with the SR 
constrains backgrounds and reduces systematic uncertainties. 
The kinematic distributions used in 
the fit are the $\chi^2$ of the kinematical reconstruction for the SR, the leading lepton's 
\pt\ for the non-prompt leptons and \ttz\ CRs, the
transverse mass for the $WZ$ CR, and the
reconstructed mass of the four leptons for the $ZZ$ CR.

The statistical analysis to extract the signal is based on a binned likelihood function 
$L(\mu,\theta)$ constructed as a product of Poisson probability terms over 
all bins in each considered distribution, and Gaussian constraint 
terms for $\theta$, a set of nuisance parameters that parameterize effects of 
statistical and systematic uncertainties on the signal and 
background expectations. The
parameter $\mu$ is a multiplicative factor for the number of signal events 
normalized to a branching ratio ${\mathcal{B}}_{\mathrm{ref}}(\tqz)=0.1\%$. 
The nuisance parameters are floated in the combined fit to adjust the expectations for signal and background 
according to the corresponding systematic uncertainties, and their fitted 
values are the adjustment that best fits the data.

The test statistic is the profile likelihood ratio
$q_{\mu}=-2\ln(L(\mu,\hat{\hat{\theta}}_{\mu})/L(\hat{\mu},\hat{\theta}))$, 
where $\hat{\mu}$ and $\hat{\theta}$ are the values of the parameters that 
maximize the likelihood function (with the constraints $0\leq \hat{\mu}\leq 
\mu$), and $\hat{\hat{\theta}}_{\mu}$ are the values of the nuisance 
parameters that maximize the likelihood function for a given value of $\mu$. 
This test statistic is used to measure the probability that the observed data is compatible
with the background-only hypothesis (i.e. for $\mu=0$) and to make 
statistical inferences about $\mu$.

\begin{figure}[htbp]
\begin{center}
  \includegraphics[width=.32\textwidth]{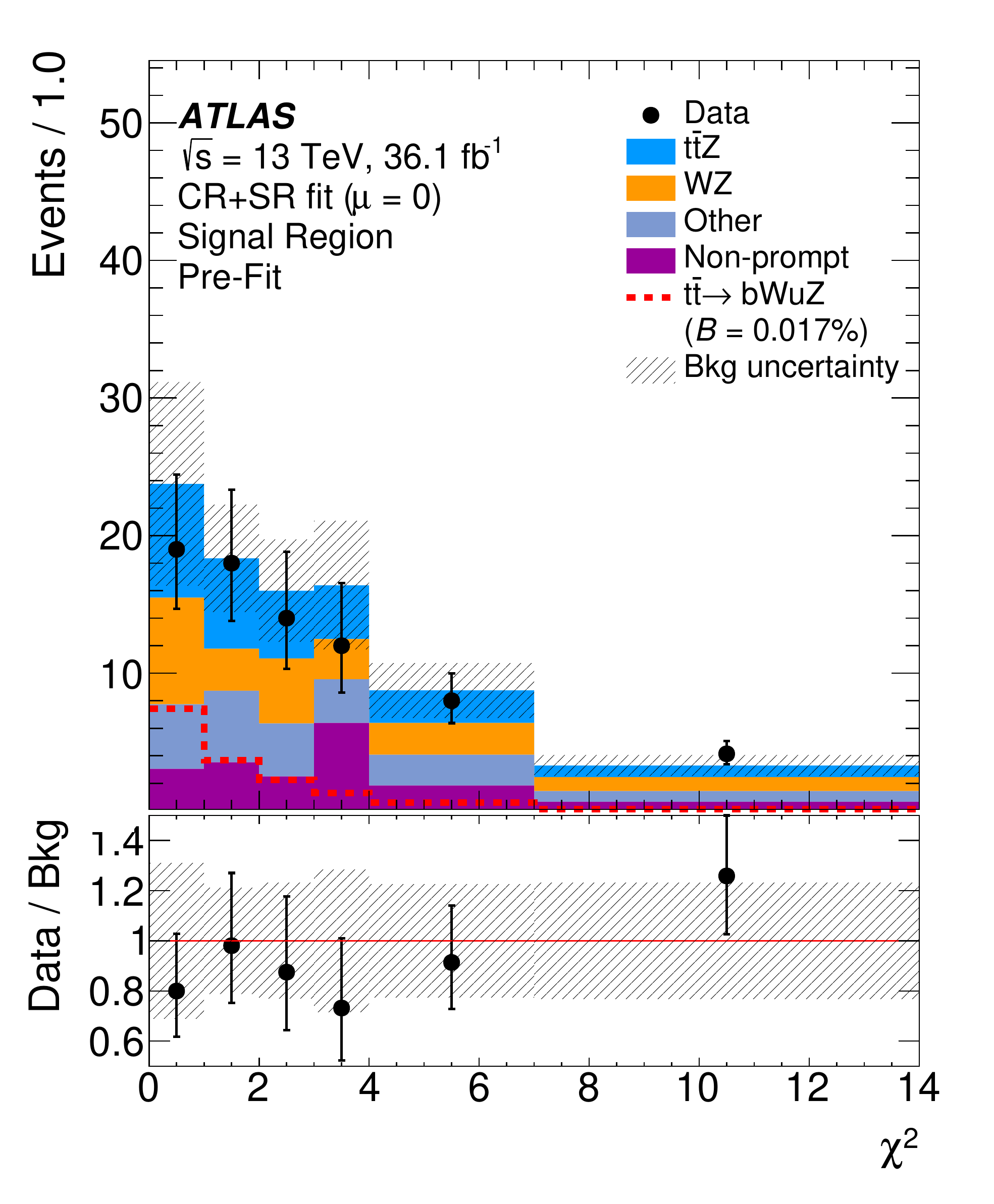}
  \includegraphics[width=.32\textwidth]{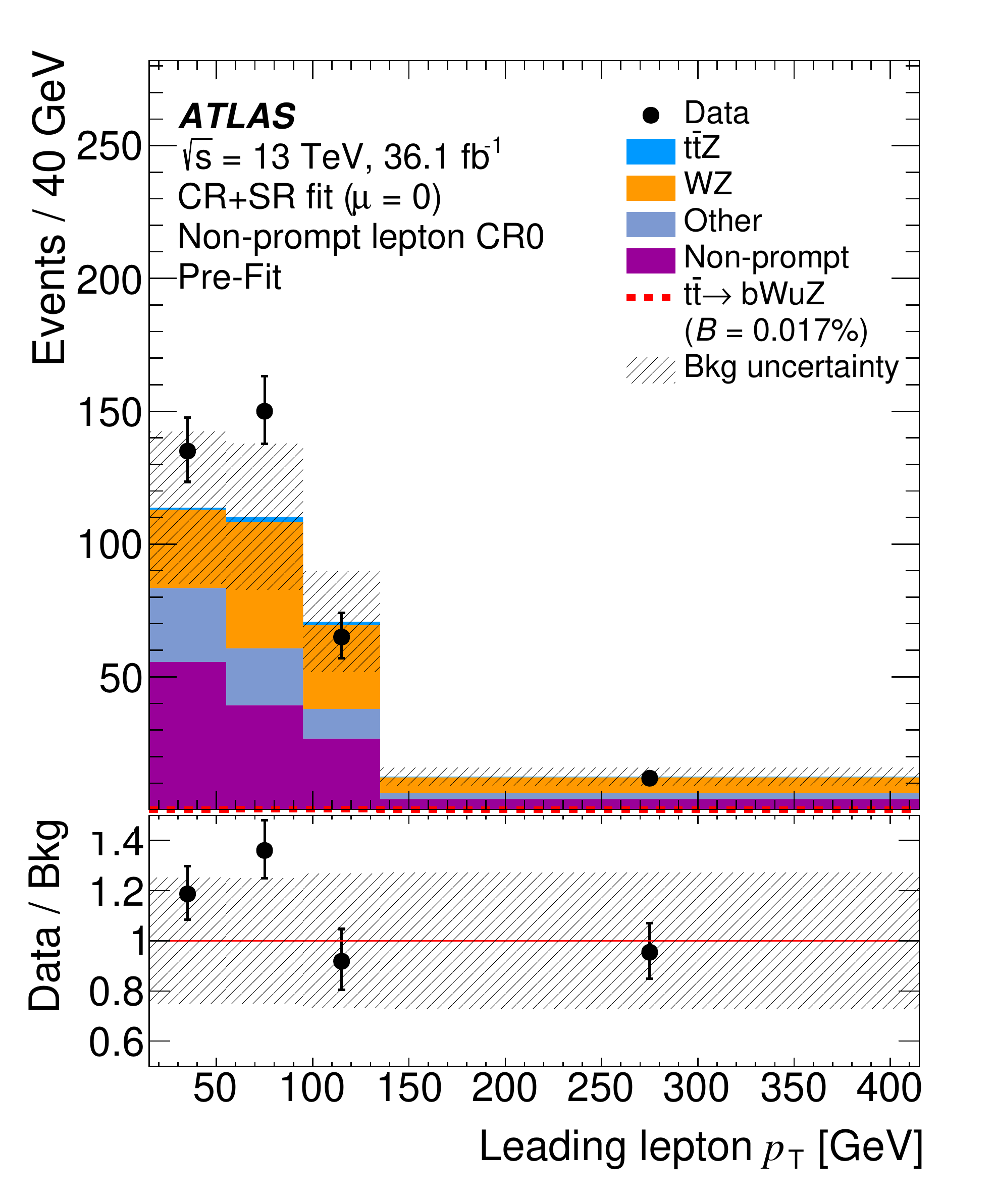}
  \includegraphics[width=.32\textwidth]{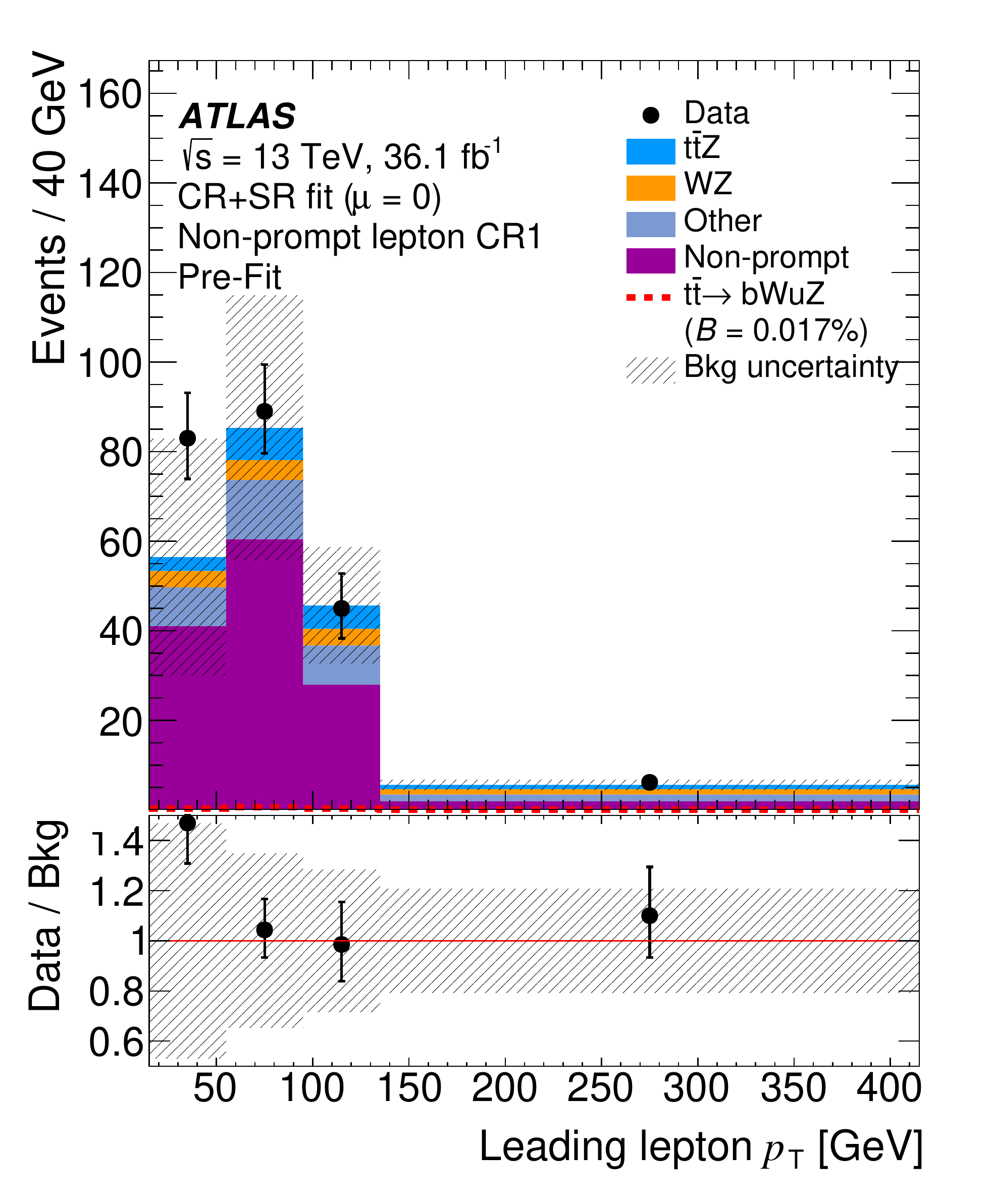}

  (a)\hspace*{.31\textwidth}(b)\hspace*{.31\textwidth}(c)

  \includegraphics[width=.32\textwidth]{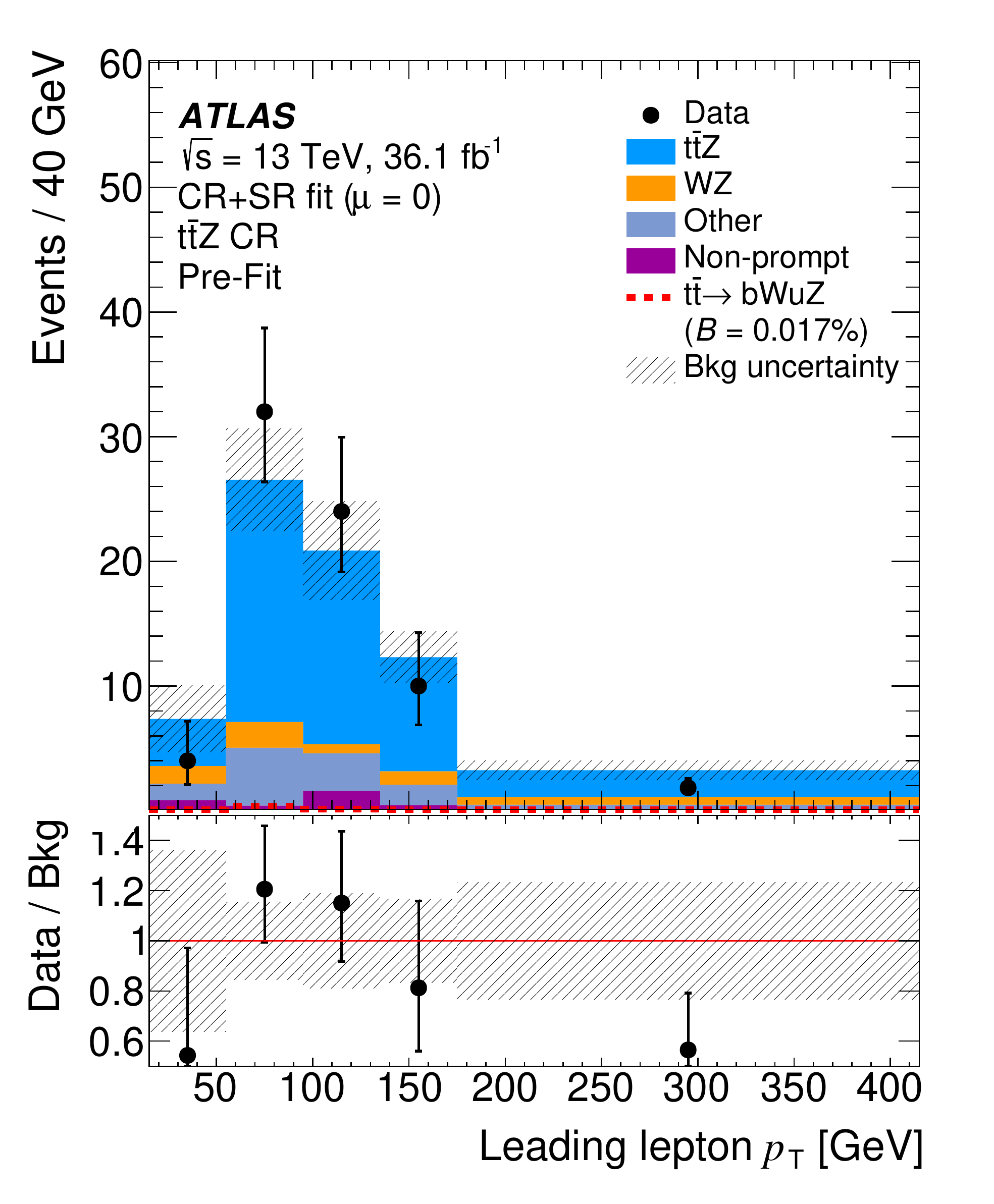}
  \includegraphics[width=.32\textwidth]{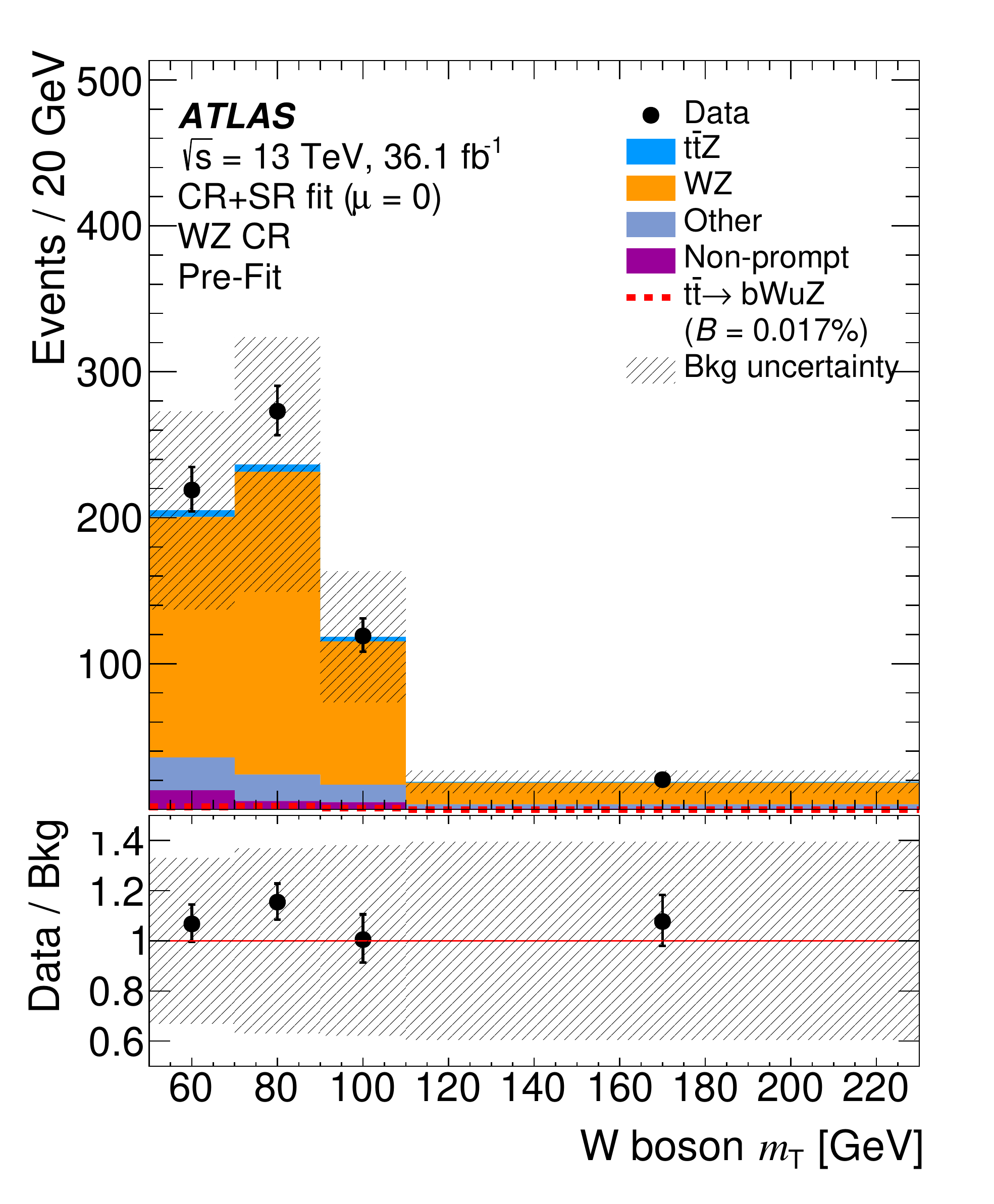}
  \includegraphics[width=.32\textwidth]{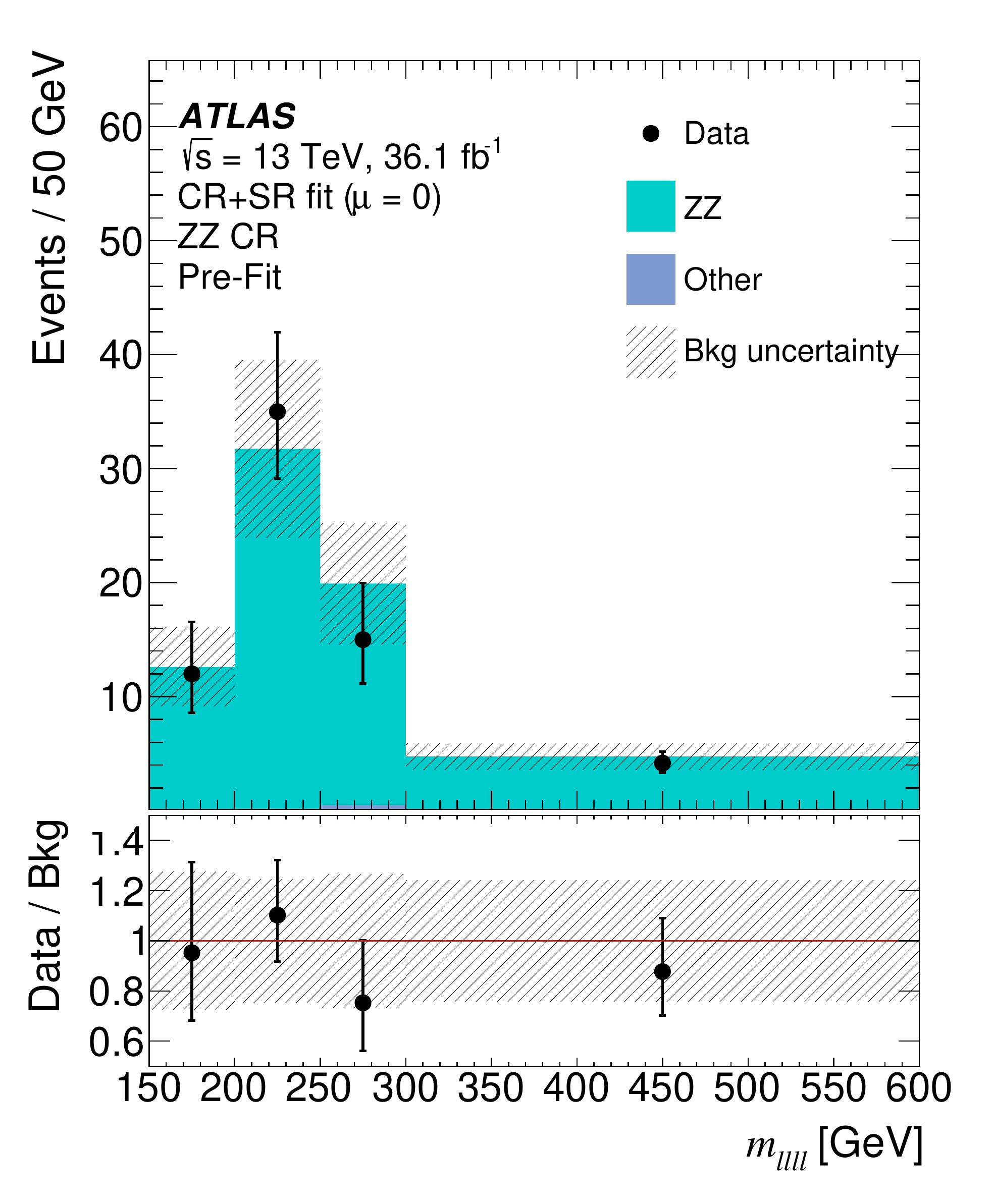}

  (d)\hspace*{.31\textwidth}(e)\hspace*{.31\textwidth}(f)
\end{center}
  \caption{Expected (filled histogram) and observed (points with error bars) distributions 
  before the combined fit under the background-only hypothesis
  of
  (a)~the $\chi^2$ of the kinematical reconstruction in the SR;
  (b)~\pt\ of the leading lepton in the non-prompt lepton CR with $b$-tag veto;
  (c)~\pt\ of the leading lepton in the non-prompt lepton CR with $b$-tag; 
  (d)~\pt\ of the leading lepton in the \ttz\ CR;
  (e)~the transverse mass in the $WZ$ CR and 
  (f)~the reconstructed mass of the four leptons in the $ZZ$ CR.
  For comparison, distributions for the FCNC $\ttbar \to bWuZ$ signal (dashed 
  line), normalized to the observed limit, are also shown.
  The ``Other'' category includes all remaining backgrounds described in 
  Section~\ref{sec:samples}.
  The dashed area represents the total uncertainty in the background prediction.
}
  \label{fig:prefit}
  \par\vspace{0pt}
\end{figure}

\begin{figure}[htbp]
\begin{center}
  \includegraphics[width=.32\textwidth]{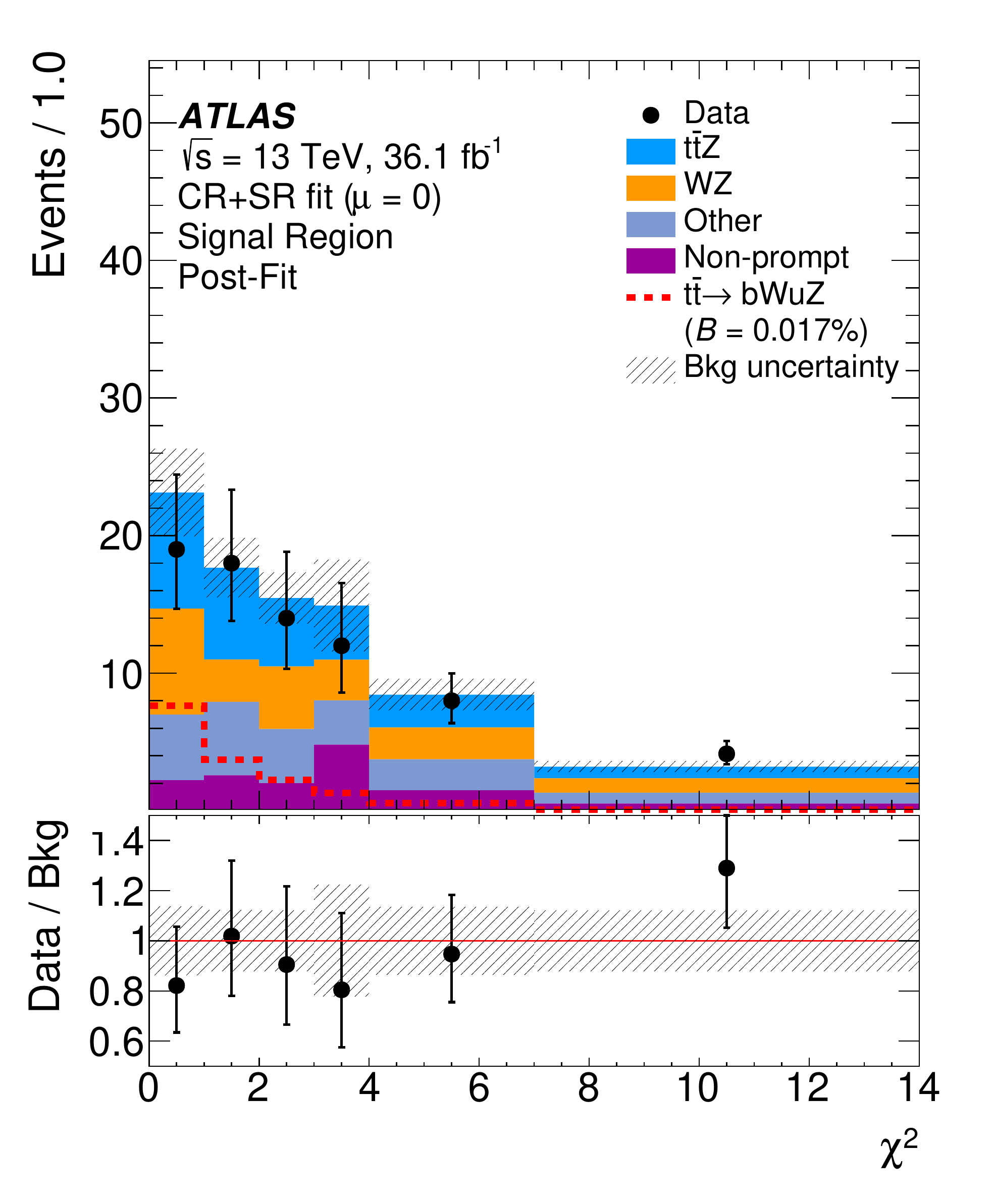}
  \includegraphics[width=.32\textwidth]{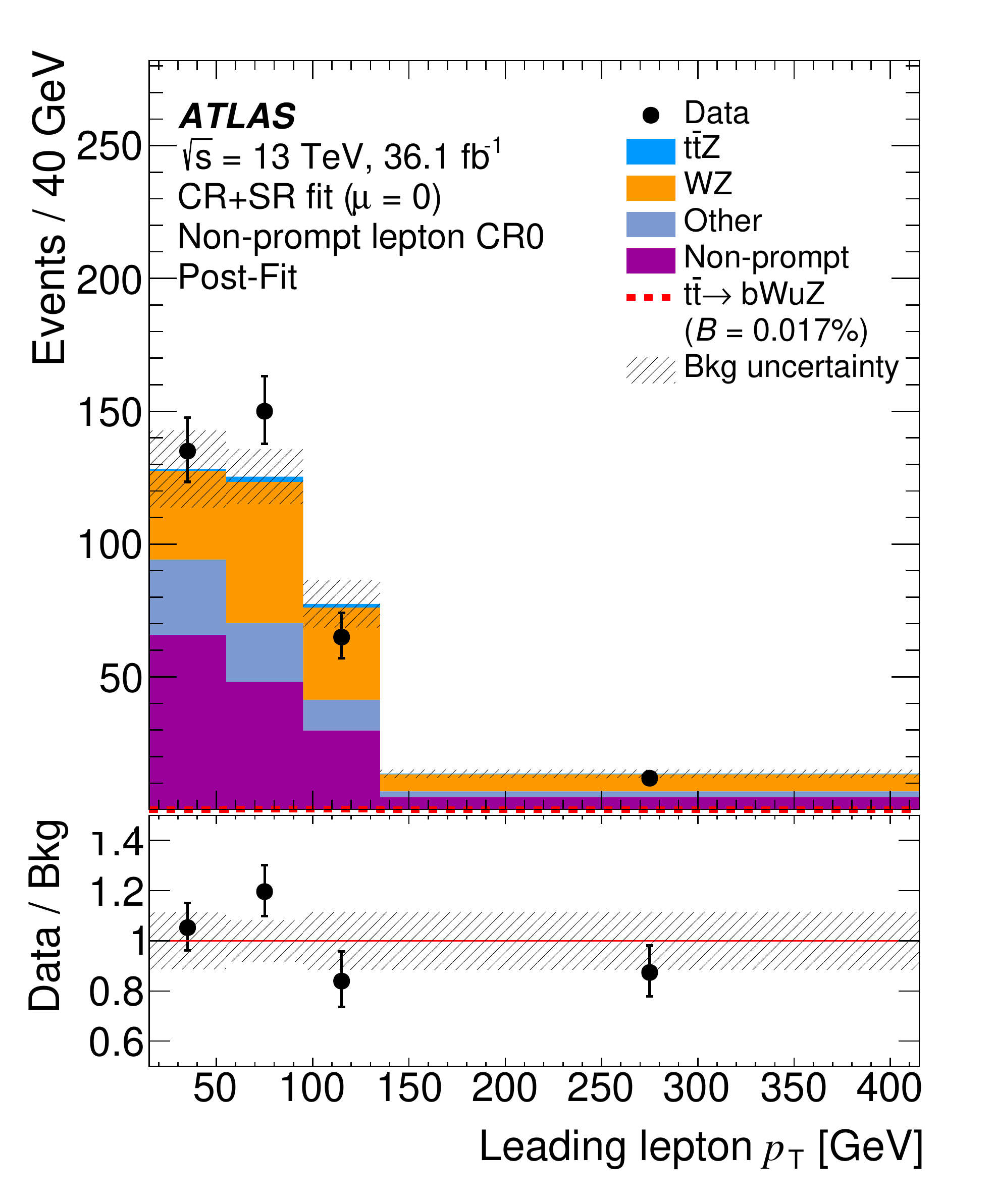}
  \includegraphics[width=.32\textwidth]{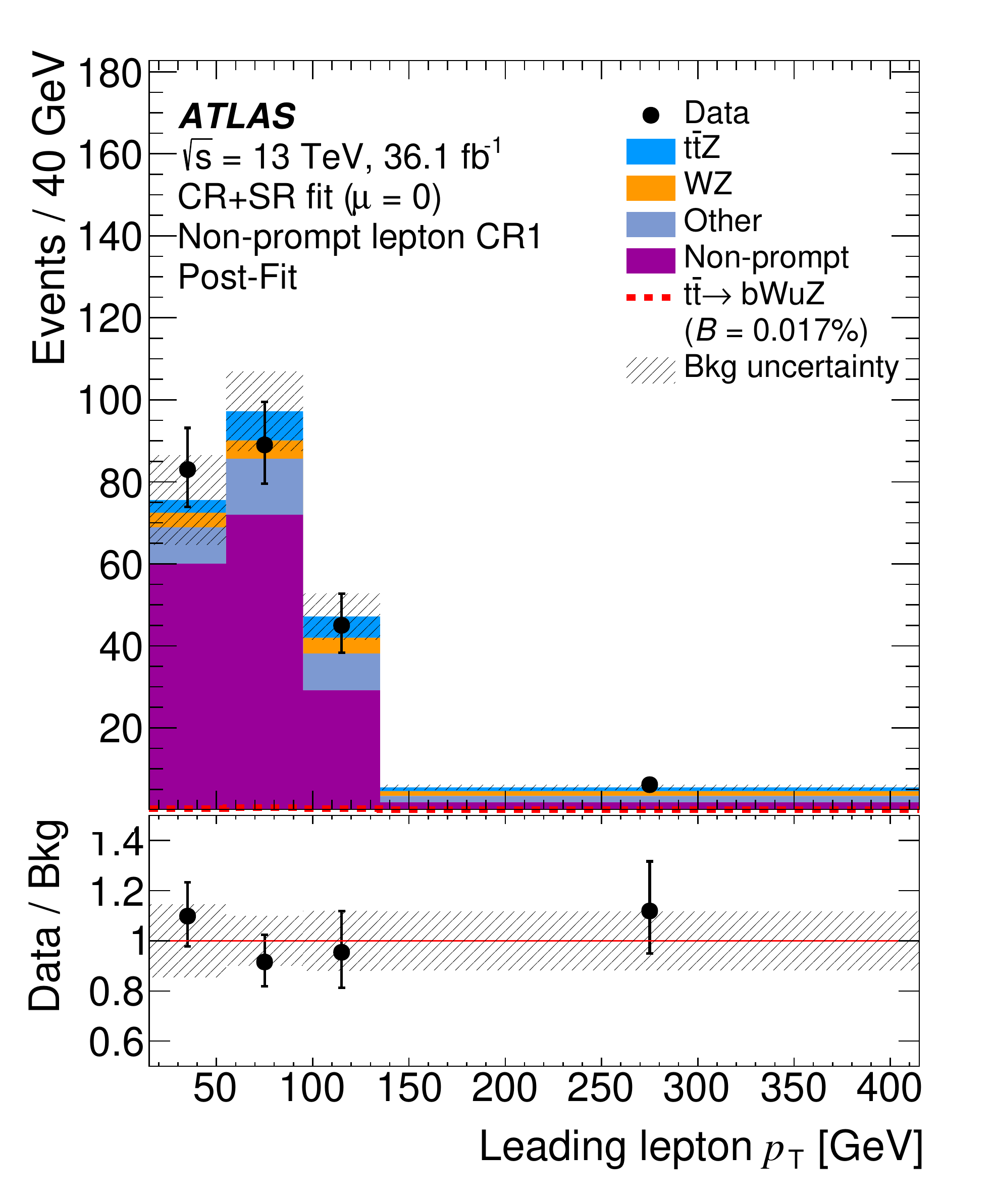}
  
  (a)\hspace*{.31\textwidth}(b)\hspace*{.31\textwidth}(c)

  \includegraphics[width=.32\textwidth]{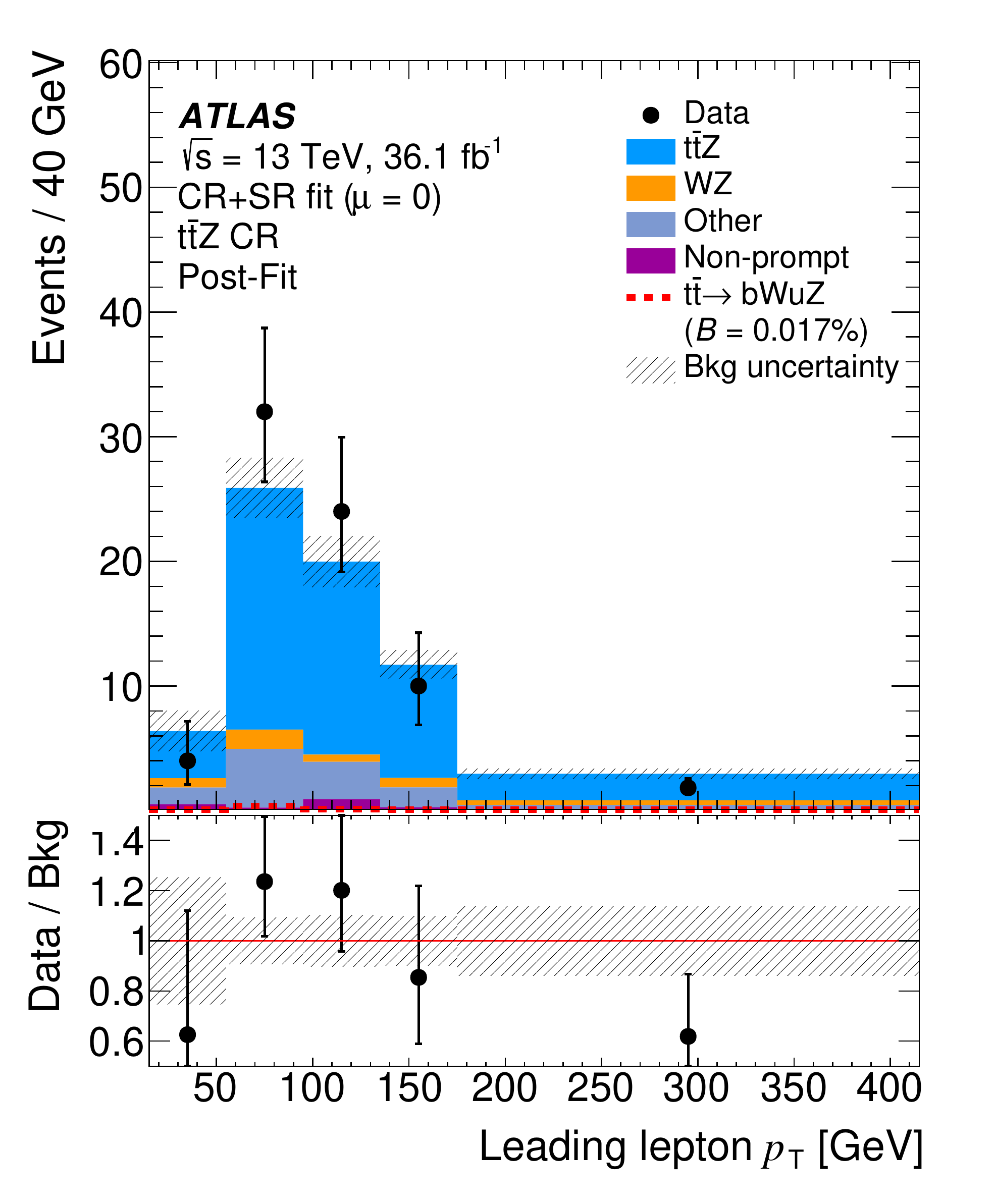}
  \includegraphics[width=.32\textwidth]{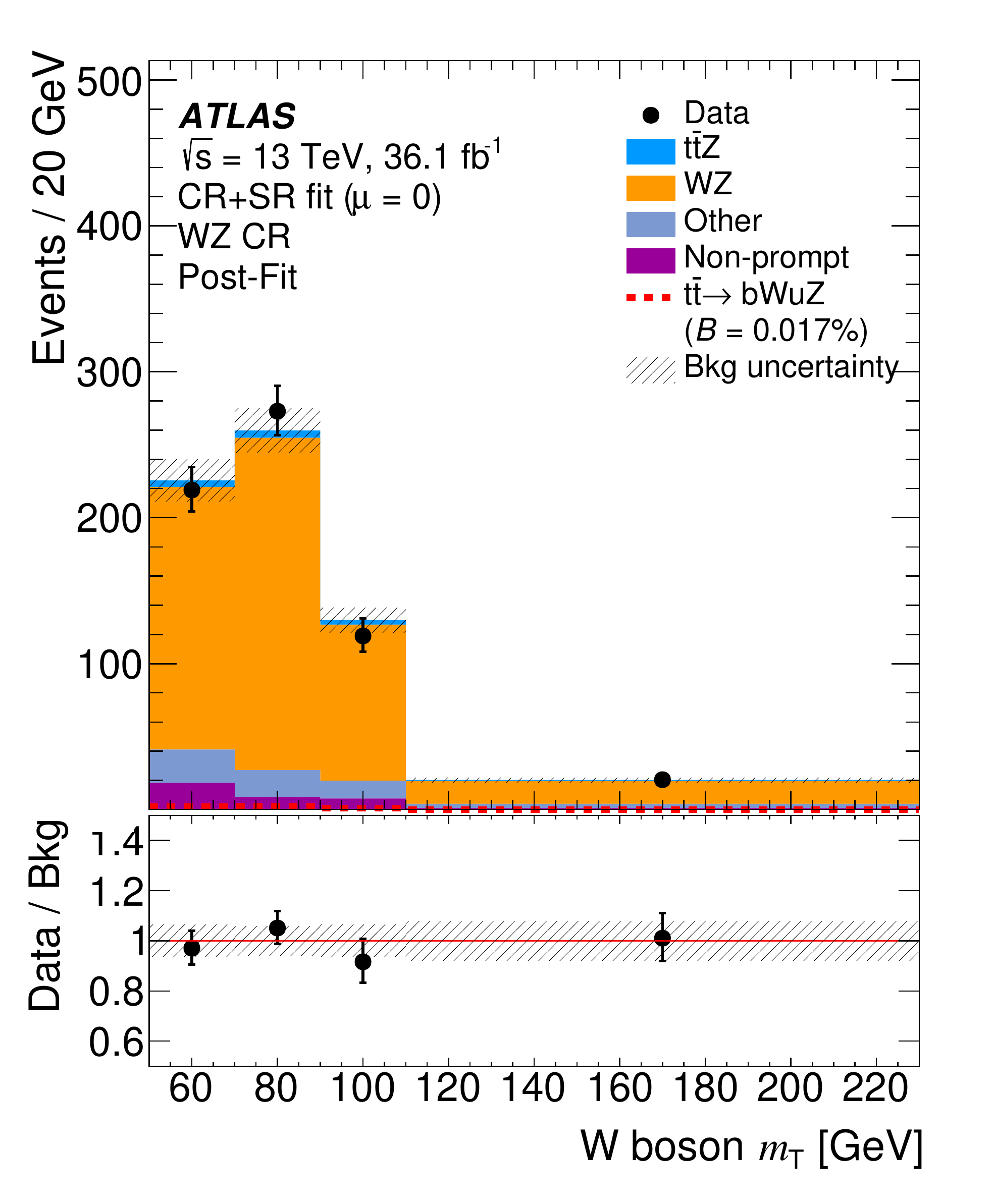}
  \includegraphics[width=.32\textwidth]{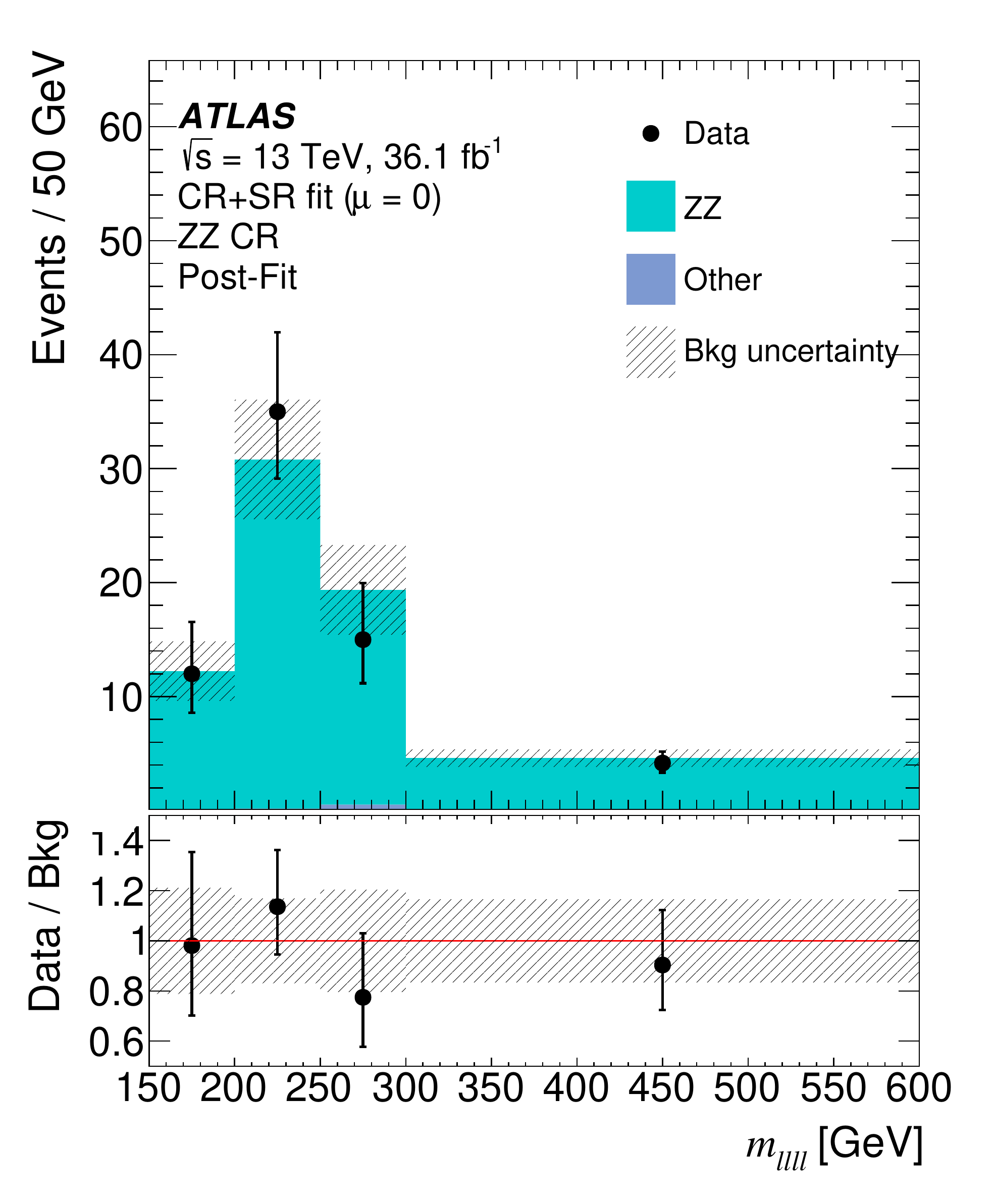}

  (d)\hspace*{.31\textwidth}(e)\hspace*{.31\textwidth}(f)
\end{center}
  \caption{Expected (filled histogram) and observed (points with error bars) distributions 
  after the combined fit under the background-only hypothesis
  of 
  (a)~the $\chi^2$ of the kinematical reconstruction in the SR;
  (b)~\pt\ of the leading lepton in the non-prompt lepton CR with $b$-tag veto;
  (c)~\pt\ of the leading lepton in the non-prompt lepton CR with $b$-tag; 
  (d)~\pt\ of the leading lepton in the \ttz\ CR;
  (e)~the transverse mass in the $WZ$ CR and 
  (f)~the reconstructed mass of the four leptons in the $ZZ$ CR.
  For comparison, distributions for the FCNC $\ttbar \to bWuZ$ signal (dashed 
  line), normalized to the observed limit, are also shown.
  The ``Other'' category includes all remaining backgrounds described in 
  Section~\ref{sec:samples}.
  The dashed area represents the total uncertainty in the background prediction.
}
  \label{fig:postfit}
  \par\vspace{0pt}
\end{figure}

\begin{figure}[htbp]
\begin{center}
  \includegraphics[width=.49\textwidth]{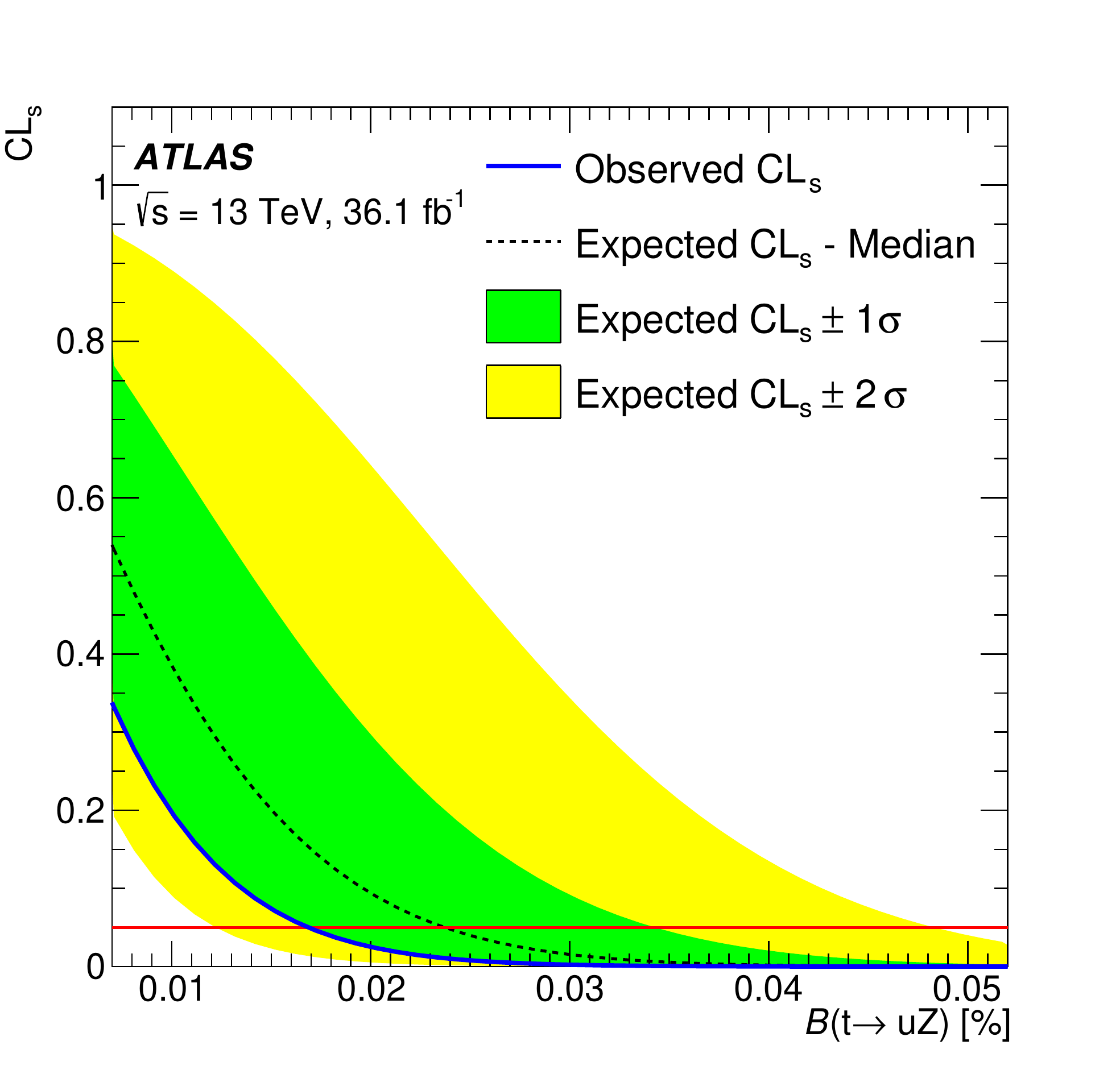} 
  \includegraphics[width=.49\textwidth]{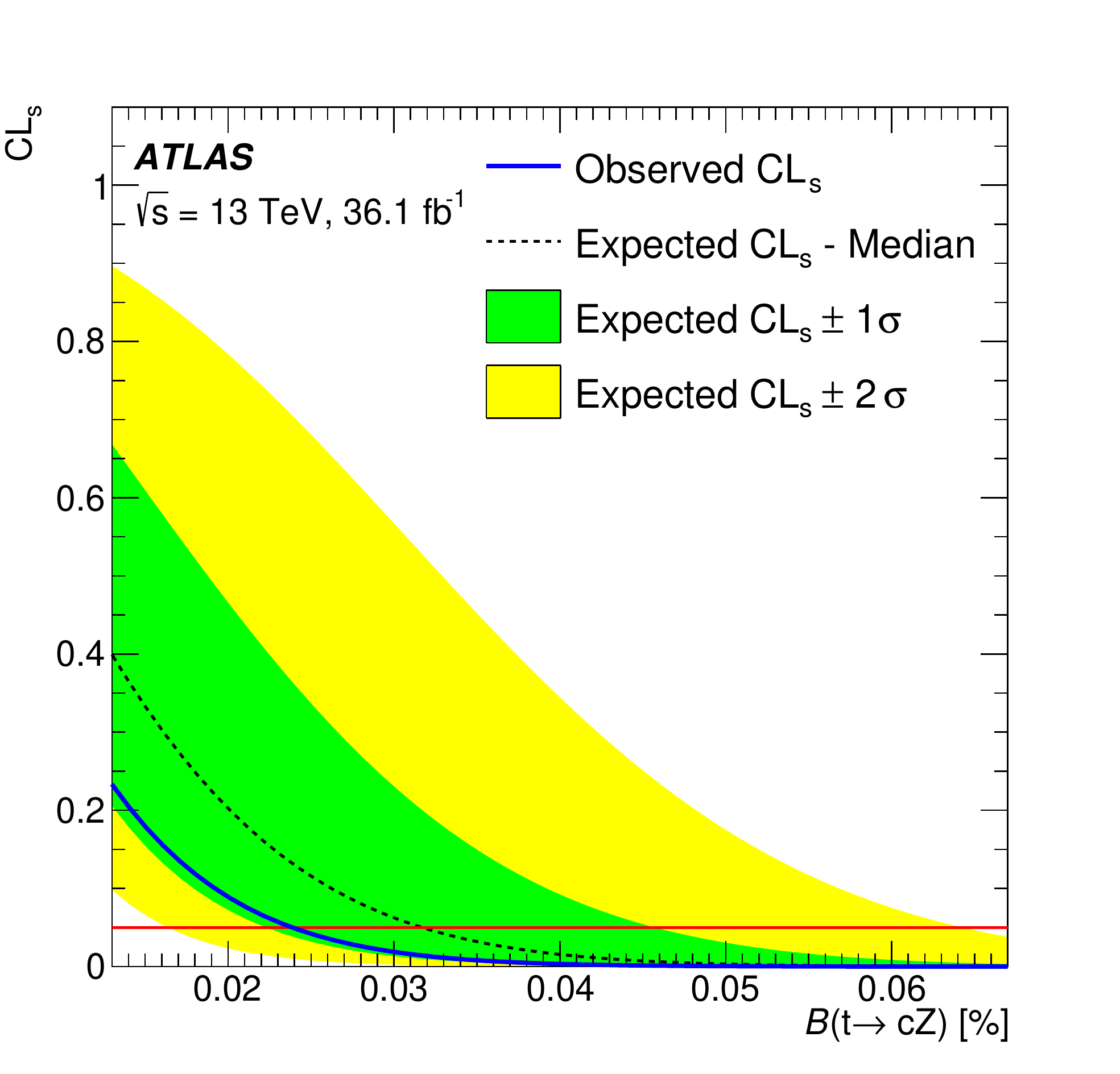} 

  (a)\hspace*{.48\textwidth}(b)
\end{center}
\caption{(a) CL$_\mathrm{s}$ vs ${\mathcal{B}}(\tuz)$ and (b) CL$_\mathrm{s}$ vs ${\mathcal{B}}(\tcz)$ taking into 
	account systematic and statistical uncertainties. The  observed  CL$_\mathrm{s}$ values
	(solid line) are compared to the expected (median)  CL$_\mathrm{s}$ 
	values under the background-only hypothesis (dashed line). The surrounding shaded
	bands correspond to the 68\% and 95\% CL intervals around the expected CL$_\mathrm{s}$
    values, denoted by $\pm1\sigma$ and $\pm2\sigma$, respectively.
    The solid line at CL$_\mathrm{s}$ = 0.05 denotes the threshold below 
	which the hypothesis is excluded at 95\% CL.}
\label{fig:result-cls}
\end{figure}

The distributions used in the combined fit under the background-only hypothesis are presented in
Figure~\ref{fig:prefit}. The same distributions after the fit are presented in
Figure~\ref{fig:postfit}.
Table~\ref{tab:signal} shows the expected number of background events, number 
of selected data events, and signal yields in the SR before and after the fit.
The post-fit signal yield changes relative to the pre-fit one
due to the fitted nuisance parameters.
The yields in the CRs after the fit are shown in
Table~\ref{tab:backgrounds:afterfit}.
Good agreement between data and the expectation from the background-only hypothesis
is observed, and no evidence of a FCNC signal is found. 
The upper limits on ${\mathcal{B}}(t\to qZ)$ are computed with the CL$_\mathrm{s}$
method~\cite{Read:2002hq, Junk:1999kv} using the asymptotic properties of 
$q_{\mu}$~\cite{Cowan:2010js, Verkerke:2003ir, Moneta:2010pm} and assuming that only one FCNC mode contributes. 
Figure~\ref{fig:result-cls} shows the observed CL$_\mathrm{s}$ for ${\mathcal{B}}(\tuz)$ and ${\mathcal{B}}(\tcz)$
together with the $\pm1\sigma$ and $\pm2\sigma$ bands
for the expected values. The 95\% confidence level (CL) limit on 
${\mathcal{B}}(t\to uZ)$ is $1.7\times10^{-4}$, and on ${\mathcal{B}}(t\to cZ)$ it is $2.4\times10^{-4}$.
The observed and expected limits are shown in Table~\ref{tab:limits}. 
It can be seen that the observed limit is about $1\sigma$ more stringent than 
expected due to a smaller number of observed events in the first bin of the 
$\chi^2$ distribution in the SR, which is the one with the largest sensitivity 
to the signal.

Using the effective field theory framework developed in the 
TopFCNC model~\cite{Degrande:2014tta,Durieux:2014xla} and
assuming a cut-off scale $\Lambda=1~\TeV$ and that only one operator has a 
non-zero value, the upper limits on ${\mathcal{B}}(t\to uZ)$ and ${\mathcal{B}}(t\to cZ)$ are
converted to 95\% CL upper limits on the moduli of the operators
contributing to the FCNC decay \tqz, which are presented in Table~\ref{tab:unblinded:operators}.

\begin{table}[htbp]

\caption{ Expected number of background events, number of selected data
events, and number of signal events (arbitrarily normalized to a branching ratio of 
$\mathcal{B}$(\tqz) = 0.1\%) in the signal region before and after the combined fit under the background-only hypothesis. 
The uncertainties shown include all of the systematic uncertainties described in Section~\ref{sec:systematics}. The entry labelled 
``other backgrounds'' includes all remaining backgrounds described in 
Section~\ref{sec:samples} and in Table~\ref{tab:samples}.
The uncertainties in the post-fit yields are calculated using the full correlation matrix from the fit.}
\label{tab:signal}
\begin{center}
	\begin{tabular}{lr@{\,}@{$\pm$}@{\,}lr@{\,}@{$\pm$}@{\,}l}
		\hline
		Sample			& \multicolumn{4}{c}{Yields}					\\
		& \multicolumn{2}{c}{Pre-fit} & \multicolumn{2}{c}{Post-fit}			\\
		\hline
		\ttz & \num[round-mode=figures,round-precision=2]{36.6997} & \num[round-mode=figures,round-precision=1]{5.0201} 	& 
		\num[round-mode=figures,round-precision=2]{37.1146} & \num[round-mode=figures,round-precision=1]{4.2905}		\\
		$WZ$	 & \num[round-mode=figures,round-precision=2]{32.3933} & \num[round-mode=figures,round-precision=2]{18.7714}	& 
		\num[round-mode=figures,round-precision=2]{32.4839} & \num[round-mode=figures,round-precision=1]{7.66832} 	\\
		$ZZ$	 & \num[round-mode=figures,round-precision=2]{6.20986} & \num[round-mode=figures,round-precision=2]{3.18103}	& 
		\num[round-mode=figures,round-precision=2]{6.44883} & \num[round-mode=figures,round-precision=2]{2.97054} \\
		Non-prompt leptons & \num[round-mode=figures,round-precision=2]{25.6236} & \num[round-mode=figures,round-precision=2]{10.8477}
		& \num[round-mode=figures,round-precision=2]{19.828} & \num[round-mode=figures,round-precision=1]{6.90295} 	\\
		Other backgrounds  & \num[round-mode=figures,round-precision=2]{22.8885} & \num[round-mode=figures,round-precision=1]{4.28099}	&
		\num[round-mode=figures,round-precision=2]{23.1155} & \num[round-mode=figures,round-precision=1]{4.19727} \\
		\hline
		Total background	 & \num[round-mode=figures,round-precision=3]{124.117} & \num[round-mode=figures,round-precision=2]{25.7469}& 
		\num[round-mode=figures,round-precision=3]{118.991} & \num[round-mode=figures,round-precision=2]{10.00}	\\
		\hline
		Data	& \multicolumn{2}{c}{116}	& \multicolumn{2}{c}{116}	\\
		\hline
		Data / Bkg   & 	\num[round-mode=figures,round-precision=2]{0.936882} & \num[round-mode=figures,round-precision=2]{0.212881}
		&
		\num[round-mode=figures,round-precision=2]{0.974865} & \num[round-mode=figures,round-precision=2]{0.121775}  \\
		\hline
		Signal \tuz\ ($\mathcal{B}$ = 0.1\%) & \num[round-mode=figures,round-precision=3]{101.4958} & \num[round-mode=figures,round-precision=1]{7.94576}	&  
		\num[round-mode=figures,round-precision=3]{102.655} & \num[round-mode=figures,round-precision=1]{7.55106}\\
		Signal \tcz\ ($\mathcal{B}$ = 0.1\%) & \num[round-mode=figures,round-precision=2]{85.4815} & \num[round-mode=figures,round-precision=1]{6.90468} &
		\num[round-mode=figures,round-precision=2]{86.6599} & \num[round-mode=figures,round-precision=1]{6.61046}	\\
		\hline
	\end{tabular}
\end{center}
\end{table}

\begin{table}[htbp]

\caption{Event yields in the background control regions for all significant 
sources of events  
after the combined fit under the background-only hypothesis. 
The uncertainties shown include all of the systematic uncertainties described in Section~\ref{sec:systematics}. 
The entry labelled ``other backgrounds'' includes all remaining backgrounds described in Section~\ref{sec:samples} 
and in Table~\ref{tab:samples}.
The uncertainties in the post-fit yields are calculated using the 
full correlation matrix from the fit.}
\label{tab:backgrounds:afterfit}
\begin{center}
	\begin{tabular}{lr@{\,}@{$\pm$}@{\,}lr@{\,}@{$\pm$}@{\,}lr@{\,}@{$\pm$}@{\,}lr@{\,}@{$\pm$}@{\,}lr@{\,}@{$\pm$}@{\,}l}
		\hline
		Sample  	& \multicolumn{2}{c}{\ttz\ CR} & \multicolumn{2}{c}{$WZ$ CR} & \multicolumn{2}{c}{$ZZ$ CR}	& \multicolumn{2}{c}{Non-prompt} 		& \multicolumn{2}{c}{Non-prompt}	\\
		&\multicolumn{2}{c}{}&\multicolumn{2}{c}{}&\multicolumn{2}{c}{}&  \multicolumn{2}{c}{lepton CR0} & \multicolumn{2}{c}{lepton CR1} \\
		\hline
		\ttz & \num[round-mode=figures,round-precision=2]{60.5218} & \num[round-mode=figures,round-precision=1]{6.40399} & 
		\num[round-mode=figures,round-precision=3]{16.4627} & \num[round-mode=figures,round-precision=2]{3.05579} &
		\num[round-mode=figures,round-precision=2]{0} & \num[round-mode=figures,round-precision=2]{0} & \num[round-mode=figures,round-precision=2]{6.13246} & \num[round-mode=figures,round-precision=2]{1.21101} & 
		\num[round-mode=figures,round-precision=3]{21.8991} & \num[round-mode=figures,round-precision=2]{2.9037} \\
		$WZ$ & \num[round-mode=figures,round-precision=1]{5.83642} & \num[round-mode=figures,round-precision=1]{3.97495} & 
		\num[round-mode=figures,round-precision=2]{608.315} & \num[round-mode=figures,round-precision=1]{38.7664} & 
		\num[round-mode=figures,round-precision=2]{0} & \num[round-mode=figures,round-precision=2]{0} & \num[round-mode=figures,round-precision=3]{166.01} & \num[round-mode=figures,round-precision=2]{13.1637} & 
		\num[round-mode=figures,round-precision=2]{20.0935} & \num[round-mode=figures,round-precision=1]{5.02067} \\ 
		$ZZ$ & \num[round-mode=figures,round-precision=1]{0.0683511} & \num[round-mode=figures,round-precision=1]{0.0226991} & \num[round-mode=figures,round-precision=2]{48.5672} & \num[round-mode=figures,round-precision=1]{9.06984} & 
		\num[round-mode=figures,round-precision=2]{89.0181} & \num[round-mode=figures,round-precision=2]{11.5585} & 
		\num[round-mode=figures,round-precision=2]{58.6857} & \num[round-mode=figures,round-precision=2]{10.2439} & 
		\num[round-mode=figures,round-precision=2]{8.95639} & \num[round-mode=figures,round-precision=2]{2.2026} \\
		Non-prompt leptons & \num[round-mode=figures,round-precision=2]{1.95248} & \num[round-mode=figures,round-precision=2]{2.31693} & 
		\num[round-mode=figures,round-precision=2]{40.9502} & \num[round-mode=figures,round-precision=2]{14.7821} & 
		\num[round-mode=figures,round-precision=2]{0} & \num[round-mode=figures,round-precision=2]{0} & \num[round-mode=figures,round-precision=3]{176.602} & \num[round-mode=figures,round-precision=2]{32.2561} & 
		\num[round-mode=figures,round-precision=3]{174.241} & \num[round-mode=figures,round-precision=2]{20.9713} \\
		Other backgrounds & \num[round-mode=figures,round-precision=3]{13.3662} & \num[round-mode=figures,round-precision=2]{2.62167} & 
		\num[round-mode=figures,round-precision=2]{22.683} & \num[round-mode=figures,round-precision=1]{5.43999} & 
		\num[round-mode=figures,round-precision=2]{1.05085} & \num[round-mode=figures,round-precision=1]{0.612449} & 
		\num[round-mode=figures,round-precision=2]{18.6577} & \num[round-mode=figures,round-precision=1]{6.36586} & 
		\num[round-mode=figures,round-precision=2]{33.1435} & \num[round-mode=figures,round-precision=1]{6.58838} \\
		\hline 
		Total background & \num[round-mode=figures,round-precision=2]{81.7453} & \num[round-mode=figures,round-precision=1]{7.11639} & 
		\num[round-mode=figures,round-precision=3]{736.978} & \num[round-mode=figures,round-precision=2]{34.6995} & 
		\num[round-mode=figures,round-precision=2]{90.0689} & \num[round-mode=figures,round-precision=2]{11.5543} & 
		\num[round-mode=figures,round-precision=3]{426.088} & \num[round-mode=figures,round-precision=2]{30.1122} & 
		\num[round-mode=figures,round-precision=3]{258.334} & \num[round-mode=figures,round-precision=2]{20.1599} \\
		\hline 
		Data   & \multicolumn{2}{c}{81} & \multicolumn{2}{c}{734} & \multicolumn{2}{c}{87} & \multicolumn{2}{c}{433\pho} & \multicolumn{2}{c}{260\pho}  \\ 
		\hline 
		Data / Bkg   & \num[round-mode=figures,round-precision=2]{0.990883} & \num[round-mode=figures,round-precision=2]{0.139867} & \num[round-mode=figures,round-precision=2]{0.995959} & \num[round-mode=figures,round-precision=1]{0.059585} & \num[round-mode=figures,round-precision=2]{0.965927} & \num[round-mode=figures,round-precision=2]{0.161488} & 
		\num[round-mode=figures,round-precision=3]{1.01622} & \num[round-mode=figures,round-precision=1]{0.0868491} & \num[round-mode=figures,round-precision=3]{1.00645} & \num[round-mode=figures,round-precision=2]{0.100323} \\ 
		\hline
	\end{tabular}
\end{center}
\end{table}

\begin{table}[htbp]
\caption[95\% CL limits]{Observed and expected 95\% CL upper limits on the FCNC 
top-quark decay branching ratios. The expected 
central value is shown together with the $\pm1\sigma$ bands, which 
includes the contribution from the statistical and systematic uncertainties.}
\label{tab:limits}

\begin{center}
\begin{tabular}{lrr}
\hline
			& ${\mathcal{B}}(\tuz)$	& ${\mathcal{B}}(\tcz)$ \\
\hline
Observed		& \uZObs	& \cZObs \\
Expected $-1\sigma$	& \uZExpMinus	& \cZExpMinus \\
Expected		& \uZExp	& \cZExp \\
Expected $+1\sigma$	& \uZExpPlus	& \cZExpPlus \\
\hline
\end{tabular}
\end{center}

\end{table}

\begin{table}
	\caption{Observed and expected 95\% CL upper limits on the moduli of the operators 
	contributing to the FCNC decays $t\to uZ$ and $t\to cZ$  within the TopFCNC model 
	for a new-physics energy scale $\Lambda=1~\TeV$.  } 
	\label{tab:unblinded:operators}
	\begin{center}
	\begin{tabular}{lcc}
		\hline 
		Operator & Observed  & Expected  \\
		\hline 						
		$|C_{uB}^{(31)}|$\rule[-1.8ex]{0pt}{5.0ex} & 0.25 & 0.30 \\
		$|C_{uW}^{(31)}|$\rule[-1.8ex]{0pt}{4.8ex} & 0.25 & 0.30 \\
		$|C_{uB}^{(32)}|$\rule[-1.8ex]{0pt}{4.8ex} & 0.30 & 0.34 \\
		$|C_{uW}^{(32)}|$\rule[-1.8ex]{0pt}{4.8ex} & 0.30 & 0.34\\
		\hline 
	\end{tabular}
    \end{center}
\end{table} 

\clearpage

%% file: qz13tev-conclusions.tex
\FloatBarrier
\section{Conclusions\label{sec:conclusions}}

An analysis is performed to search for \ttbar events with one top quark decaying through the FCNC \tqz\ 
$(q = u, c)$ channel and the other through the dominant 
Standard Model mode $t\to bW$, where only $Z$ boson decays
into charged leptons and leptonic $W$ boson decays are considered as 
signal. The data were collected by the ATLAS experiment in $pp$ collisions corresponding
to an integrated luminosity of \luminosity at the LHC at a centre-of-mass energy of 
$\sqrt{s}=13~\TeV$. There is good agreement between the data and Standard Model expectations, and
no evidence of a signal is found. The 95\% CL limits on the $t\to qZ$ branching ratio
are ${\mathcal{B}}(\tuz) < \uZObs$ and ${\mathcal{B}}(\tcz) < \cZObs$,
improving previous ATLAS results by more than 60\%.
These limits constrain the values of effective field theory operators contributing 
to the $\tuz$ and $\tcz$ FCNC decays of the top quark.

%% file: acknowledgements/Acknowledgements.tex

We thank CERN for the very successful operation of the LHC, as well as the
support staff from our institutions without whom ATLAS could not be
operated efficiently.

We acknowledge the support of ANPCyT, Argentina; YerPhI, Armenia; ARC, Australia; BMWFW and FWF, Austria; ANAS, Azerbaijan; SSTC, Belarus; CNPq and FAPESP, Brazil; NSERC, NRC and CFI, Canada; CERN; CONICYT, Chile; CAS, MOST and NSFC, China; COLCIENCIAS, Colombia; MSMT CR, MPO CR and VSC CR, Czech Republic; DNRF and DNSRC, Denmark; IN2P3-CNRS, CEA-DRF/IRFU, France; SRNSFG, Georgia; BMBF, HGF, and MPG, Germany; GSRT, Greece; RGC, Hong Kong SAR, China; ISF, I-CORE and Benoziyo Center, Israel; INFN, Italy; MEXT and JSPS, Japan; CNRST, Morocco; NWO, Netherlands; RCN, Norway; MNiSW and NCN, Poland; FCT, Portugal; MNE/IFA, Romania; MES of Russia and NRC KI, Russian Federation; JINR; MESTD, Serbia; MSSR, Slovakia; ARRS and MIZ\v{S}, Slovenia; DST/NRF, South Africa; MINECO, Spain; SRC and Wallenberg Foundation, Sweden; SERI, SNSF and Cantons of Bern and Geneva, Switzerland; MOST, Taiwan; TAEK, Turkey; STFC, United Kingdom; DOE and NSF, United States of America. In addition, individual groups and members have received support from BCKDF, the Canada Council, CANARIE, CRC, Compute Canada, FQRNT, and the Ontario Innovation Trust, Canada; EPLANET, ERC, ERDF, FP7, Horizon 2020 and Marie Sk{\l}odowska-Curie Actions, European Union; Investissements d'Avenir Labex and Idex, ANR, R{\'e}gion Auvergne and Fondation Partager le Savoir, France; DFG and AvH Foundation, Germany; Herakleitos, Thales and Aristeia programmes co-financed by EU-ESF and the Greek NSRF; BSF, GIF and Minerva, Israel; BRF, Norway; CERCA Programme Generalitat de Catalunya, Generalitat Valenciana, Spain; the Royal Society and Leverhulme Trust, United Kingdom.

The crucial computing support from all WLCG partners is acknowledged gratefully, in particular from CERN, the ATLAS Tier-1 facilities at TRIUMF (Canada), NDGF (Denmark, Norway, Sweden), CC-IN2P3 (France), KIT/GridKA (Germany), INFN-CNAF (Italy), NL-T1 (Netherlands), PIC (Spain), ASGC (Taiwan), RAL (UK) and BNL (USA), the Tier-2 facilities worldwide and large non-WLCG resource providers. Major contributors of computing resources are listed in Ref.~\cite{ATL-GEN-PUB-2016-002}.

%% file: atlas_authlist.tex
 
\begin{flushleft}
{\Large The ATLAS Collaboration}

\bigskip

M.~Aaboud$^\textrm{\scriptsize 34d}$,    
G.~Aad$^\textrm{\scriptsize 99}$,    
B.~Abbott$^\textrm{\scriptsize 124}$,    
O.~Abdinov$^\textrm{\scriptsize 13,*}$,    
B.~Abeloos$^\textrm{\scriptsize 128}$,    
S.H.~Abidi$^\textrm{\scriptsize 165}$,    
O.S.~AbouZeid$^\textrm{\scriptsize 143}$,    
N.L.~Abraham$^\textrm{\scriptsize 153}$,    
H.~Abramowicz$^\textrm{\scriptsize 159}$,    
H.~Abreu$^\textrm{\scriptsize 158}$
% ...
\bigskip
\\

% ...
$^{13}$Institute of Physics, Azerbaijan Academy of Sciences, Baku; Azerbaijan.\\
% ...
$^{34}$$^{(a)}$Facult\'e des Sciences Ain Chock, R\'eseau Universitaire de Physique des Hautes Energies - Universit\'e Hassan II, Casablanca;$^{(b)}$Centre National de l'Energie des Sciences Techniques Nucleaires (CNESTEN), Rabat;$^{(c)}$Facult\'e des Sciences Semlalia, Universit\'e Cadi Ayyad, LPHEA-Marrakech;$^{(d)}$Facult\'e des Sciences, Universit\'e Mohamed Premier and LPTPM, Oujda;$^{(e)}$Facult\'e des sciences, Universit\'e Mohammed V, Rabat; Morocco.\\
% ...
$^{99}$CPPM, Aix-Marseille Universit\'e, CNRS/IN2P3, Marseille; France.\\
% ...
$^{124}$Homer L. Dodge Department of Physics and Astronomy, University of Oklahoma, Norman OK; United States of America.\\
% ...
$^{128}$LAL, Universit\'e Paris-Sud, CNRS/IN2P3, Universit\'e Paris-Saclay, Orsay; France.\\
% ...
$^{143}$Santa Cruz Institute for Particle Physics, University of California Santa Cruz, Santa Cruz CA; United States of America.\\
% ...
$^{153}$Department of Physics and Astronomy, University of Sussex, Brighton; United Kingdom.\\
% ...
$^{158}$Department of Physics, Technion, Israel Institute of Technology, Haifa; Israel.\\
$^{159}$Raymond and Beverly Sackler School of Physics and Astronomy, Tel Aviv University, Tel Aviv; Israel.\\
% ...
$^{165}$Department of Physics, University of Toronto, Toronto ON; Canada.\\
% ...
$^{a}$ Also at  Department of Physics, University of Malaya, Kuala Lumpur; Malaysia.\\
$^{b}$ Also at Borough of Manhattan Community College, City University of New York, NY; United States of America.\\
$^{c}$ Also at Centre for High Performance Computing, CSIR Campus, Rosebank, Cape Town; South Africa.\\
$^{d}$ Also at CERN, Geneva; Switzerland.\\
$^{e}$ Also at CPPM, Aix-Marseille Universit\'e, CNRS/IN2P3, Marseille; France.\\
% ...
$^{*}$ Deceased

\end{flushleft}
